\documentclass[aps,prb,10pt,twocolumn,superscriptaddress]{revtex4-2}

\usepackage{amsmath}
\usepackage{amssymb}
\usepackage{graphicx}
\usepackage{dcolumn}
\usepackage{bm}
\usepackage[utf8]{inputenc}
\usepackage{verbatim}
\usepackage{color}
\newcommand{\yrs}{YbRh$_2$Si$_2$}
\begin{document}

\title{Nuclear-order-induced quantum criticality and heavy-fermion
superconductivity at ultra-low temperatures in YbRh$_2$Si$_2$}

\author{Erwin Schuberth}
\affiliation{Physics Department, Technical University of Munich, 80333
Munich, Germany}
\author{S. Wirth}
\affiliation{Max-Planck-Institute for Chemical Physics of Solids,
N\"othnitzer Str. 40, 01187 Dresden, Germany}
\author{Frank Steglich}
\email{steglich@cpfs.mpg.de}
\affiliation{Max-Planck-Institute for Chemical Physics of Solids,
N\"othnitzer Str. 40, 01187 Dresden, Germany}
\affiliation{Center for Correlated Matter, Zhejiang University, Hangzhou,
Zhejiang 310058, China}
\date{\today}

\begin{abstract}
The tetragonal heavy-fermion metal YbRh$_2$Si$_2$ orders antiferromagnetically
at $T_{\rm N} = 70$ mK and exhibits an unconventional quantum critical point
(QCP) of Kondo-destroying type at $B_{\rm N} = 60$ mT, for the magnetic field applied within the basal ($a,b$) plane. Ultra-low-temperature magnetization
and heat-capacity measurements at very low fields indicate that the
4$f$-electronic antiferromagnetic (AF) order is strongly suppressed by a nuclear-dominated hybrid order (‘A-phase’) at $T_{\rm A} \le 2.3$ mK,
such that quantum critical fluctuations develop at $B \approx 0$
\citep{sch16}. This enables the onset of heavy-fermion superconductivity
($T_{\rm c} = 2$ mK) which appears to be suppressed by the primary AF order at
elevated temperatures. Measurements of the Meissner effect reveal bulk
superconductivity, with $T_{\rm c}$ decreasing under applied field to
$T_{\rm c} < 1$ mK at $B > 20$ mT. The observation of a weak but distinct
superconducting shielding signal at a temperature as high as 10 mK suggests
the formation of insulated random islands with emergent A-phase order and
superconductivity. Upon cooling, the shielding signal increases almost linearly
in temperature, indicating a growth of the islands which eventually percolate
at $T \approx 6.5$ mK. Recent electrical-resistivity results by \cite{ngu21} confirm the existence of superconductivity in \yrs\ at ultra-low temperatures.
The combination of the results of \cite{sch16} and \cite{ngu21} at ultra-low
temperatures below $B_{\rm N}$, along with those previously established at
higher temperatures in the paramagnetic state, provide compelling evidence
that the Kondo-destruction quantum criticality robustly drives unconventional
superconductivity.
\end{abstract}
\maketitle

\section{Kondo-destroying, field-induced quantum critical point}
\label{sec1}
Lanthanide-based intermetallic compounds showing heavy-fermion phenomena are
well understood within the framework of the Kondo lattice \citep{wir16}. The
localized open 4$f$-shells of such materials are characterized by a distinct
hierarchy of fundamental energy scales, the local Coulomb repulsion and Hund’s
rule energies, spin-orbit coupling, crystal-field splitting, Kondo screening
including excited crystal-field states at an elevated temperature, where the
formation of hybridized 4$f$-bands starts to be recognized in ARPES
measurements \citep{che17}), as well as Kondo screening of the lowest-lying
crystal-field-derived Kramers doublet ($k_{\rm B}T_{\rm K}$). As illustrated
below, the single-ion Kondo temperature $T_{\rm K}$ was found to be identical
to $T_{\rm coh}$ \citep{ern11,sei18} where spatial coherence among 4$f$-shells
sets in as seen in transport measurements.

The onsite Kondo screening competes with the inter-site magnetic
Ruderman–Kittel–Kasuya-Yosida (RKKY) interaction. While predominant Kondo
screening results in a paramagnetic heavy-fermion ground state, a dominant RKKY
interaction causes magnetic, most frequently antiferromagnetic (AF), order. For
a substantial number of these heavy-fermion metals the Kondo screening turns
out to almost exactly cancel the RKKY interaction. In this situation, a
continuous transition may exist at $T=0$ between the heavy-fermion and the AF
phase. This continuous quantum phase transition or quantum critical point (QCP)
can be tuned by a non-thermal control parameter, e.g., pressure or magnetic
field \citep{stew01,hvl07,geg08,sac11}.

\begin{figure}[t]
\begin{center}
\includegraphics[width=0.45\textwidth]{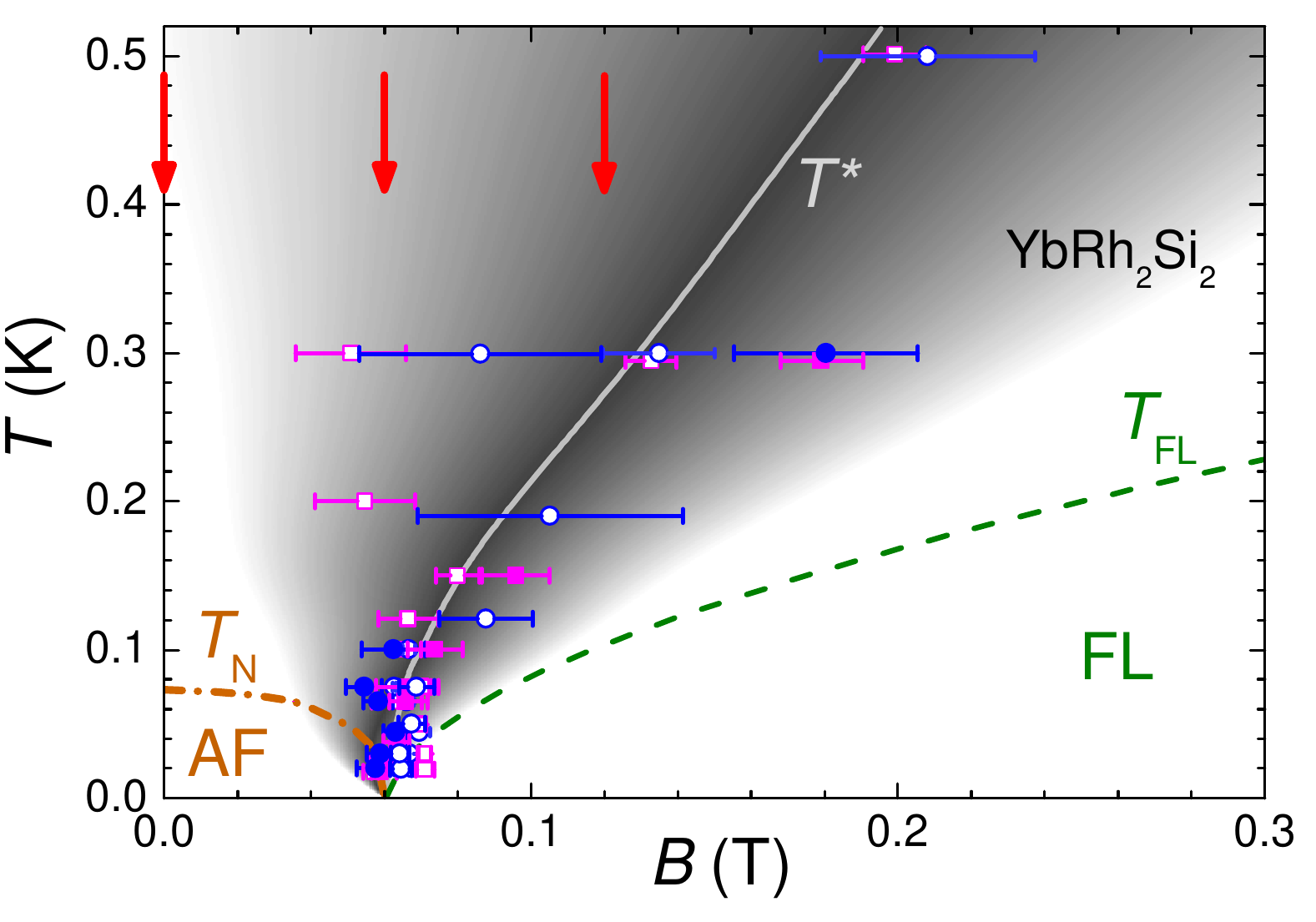}
\end{center}
\caption{Temperature--magnetic field phase diagram of \yrs\ for $B \perp c$.
The symbols mark the results from Hall and magnetoresistance measurements
on two different samples from which the Fermi surface crossover temperature
$T^*$-line is estimated \citep{Fri10}. The grey scale visualizes the slope of
isothermal magnetoresistance. Orange and green dashed lines indicate the
N\'{e}el temperature $T_{\rm N}$ and the crossover temperature $T_{\rm FL}$
below which AF order and Fermi liquid behavior are observed, respectively. Red
arrows indicate $B = 0, B_{\rm N}$ and $2B_{\rm N}$, respectively. Figure
adapted from \cite{wir16}.}
\label{phase}
\end{figure}

\yrs\ is a prototypical heavy-fermion metal \citep{ern11,koe08} which orders
antiferromagnetically at a N\'{e}el temperature $T_{\rm N} = 70$ mK
\citep{tro00}. Because of a strong magnetic anisotropy the critical field
$B_{\rm N}$ at which AF order smoothly disappears is only 60 mT when the field
is applied within the basal tetragonal ($a,b$) plane, i.e., it is more than
ten times lower compared to $B_{\rm N} \parallel c$ \citep{geg02}. The
staggered moment was shown to be very small, $\mu_{\rm AF} \approx 0.002\,
\mu_{\rm B}$ \citep{ish03}. Both the low-temperature paramagnetic ($B >
B_{\rm N}$) and AF phases ($B < B_{\rm N}$) behave as heavy Fermi liquids
\citep{geg02} whereas a funnel-shaped quantum critical regime in the $B$--$T$
phase diagram (Fig.\ \ref{phase}) with non-Fermi-liquid properties is centered
at the critical field $B_{\rm N}$. The AF phase transition is of second order
to the lowest temperature (20 mK) accessible in magnetostriction measurements
\citep{geg07}; the crossover between the paramagnetic Fermi liquid and the
non-Fermi-liquid phase turns out to be quite broad. The sub-linear white
$T^*(B)$ crossover line in Fig.\ \ref{phase} was constructed by the midpoints
of thermally broadened jumps in both the longitudinal magneto-resistivity and
the isothermally measured initial normal Hall coefficient \citep{pas04,Fri10}.
They agree satisfactorily with the locations of distinct anomalies in the
field dependence of both the isothermal DC magnetization and magnetostriction
as well as in the temperature dependence of the AC susceptibility measured at
fixed magnetic field \citep{geg07}. $T^*(B)$ indicates a crossover between a
small Fermi volume (‘small Fermi surface') at low fields and a large one at
elevated fields. The crossover width (grey shaded region) turns out to be
proportional to $T$. The QCP at $B_{\rm N} (T = 0)$ is of an unconventional
`local' (instead of itinerant) variety, which may be called a `partial Mott'
or Kondo-destroying QCP. It is characterized by a dynamical spin susceptibility
with frequency-over-temperature scaling and a fractional exponent in the
singular parts of both the frequency and temperature dependences
\citep{qsi01,col01} as observed by inelastic neutron scattering on both
UCu$_{5-x}$Pd$_x$ \citep{aro95} and CeCu$_{6-x}$Au$_x$ \citep{loe00}. The
latter material \citep{hvl94} as well as CeRhIn$_5$ \citep{shi05,par06} are
also prototypical heavy-fermion metals exhibiting a local QCP.

\begin{figure}[t]
\begin{center}
\includegraphics[width=0.48\textwidth]{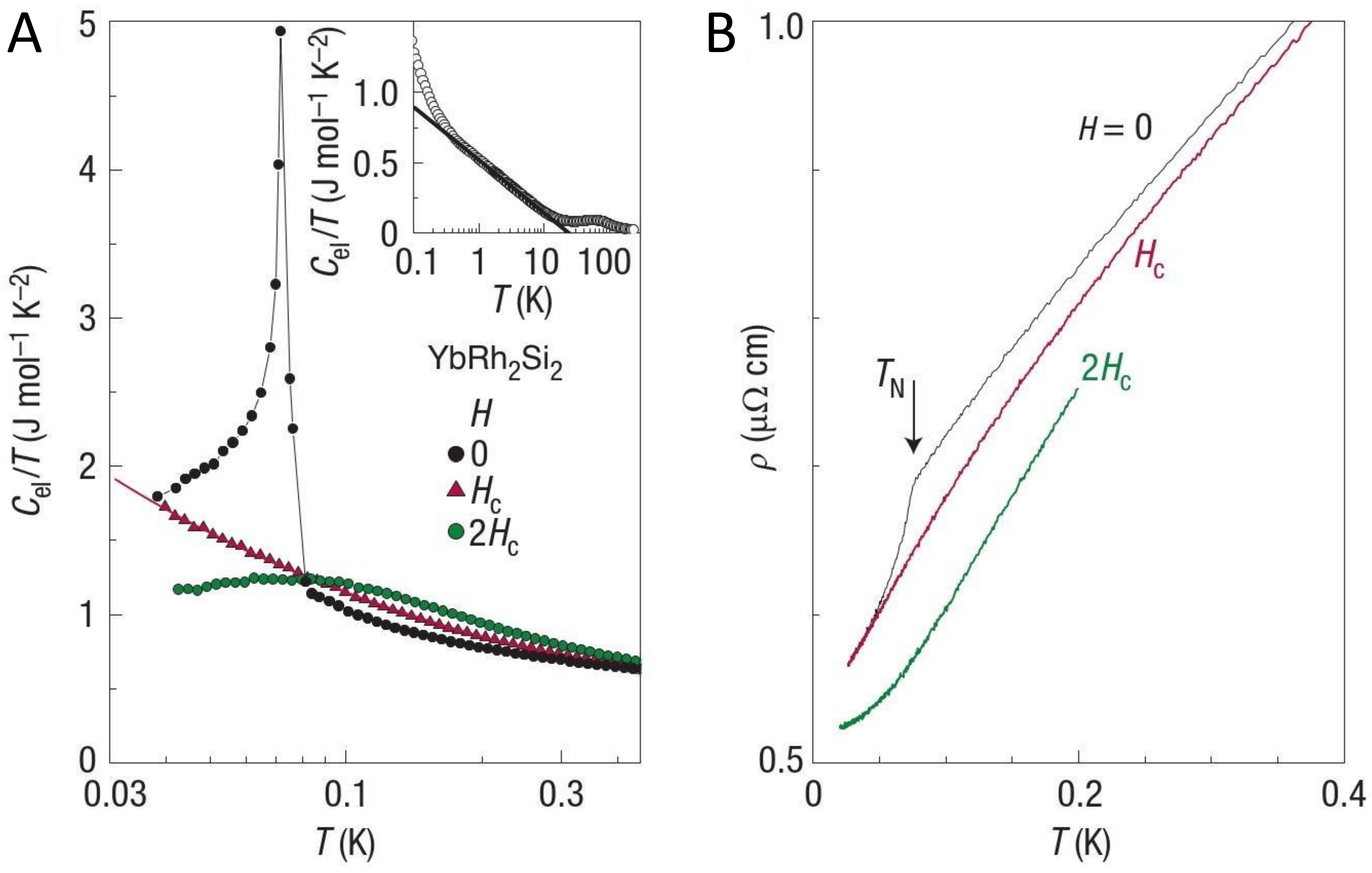}
\end{center}
\caption{\textbf{(A)} Sommerfeld coefficient of the electronic specific heat
and \textbf{(B)} electrical resistivity vs. temperature $T$ at $B = 0, B_{\rm
N}$ and $2B_{\rm N}$ for a \yrs\ single crystal with $RRR$ = 150. Inset in A
shows the Sommerfeld coefficient at $B_{\rm N} = 60$ mT between 0.1 K and room
temperature. Reproduced from \cite{geg08}.}
\label{c-rho}
\end{figure}

Figure \ref{c-rho} displays the thermal evolution at low temperatures of both
the (Sommerfeld) coefficient of the 4$f$-derived part of the specific heat,
$\gamma = C_{\rm el}/T = C_{4f}/T$ (Fig.\ \ref{c-rho}A), and the electrical
resistivity measured on an \yrs\ single crystal with a low residual
resistivity of about 0.5 $\mu$$\Omega$cm (Fig.\ \ref{c-rho}B), at $B=0$,
$B_{\rm N}$ and $2B_{\rm N}$, respectively (see red arrows in Fig.\
\ref{phase}). The behavior of a heavy Fermi liquid in both the AF and
low-temperature paramagnetic phase
is clearly resolved by a huge constant $\gamma$-value (Fig.\ \ref{c-rho}A) and
a pronounced $T^2$-dependence in $\rho(T)$ (Fig.\ \ref{c-rho}B). Very
peculiarly, the Fermi liquid in the AF phase (with small Fermi surface as
$T\rightarrow 0$) is considerably heavier than the Fermi liquid in the
paramagnetic phase (with large Fermi surface). Approaching the QCP at $B =
B_{\rm N}$ by cooling to below 0.3 K, one finds that $\gamma (T)$ diverges
following a power-law dependence with a critical exponent of $\approx - 0.3$
\citep{cus03}. As shown in the inset of Fig.\ \ref{c-rho}A, $\gamma (T)$ obeys
a logarithmic $T$-dependence \citep{cus03} at elevated temperature. The
incremental resistivity measured at $B = B_{\rm N}$ on a very clean single
crystal is strictly linear in $T$, $\Delta \rho(T) = \rho(T) - \rho_0 =
A'T$, at $T < 0.1$ K. Over a broad temperature range up to $T = 10$~K, the
resistivity semi-quantitatively follows a $T$-linear dependence, although the
`temperature-dependent resistivity exponent', defined as $ d \ln (\rho -
\rho_0)/d \ln T$, shows some variation that is clustered around $1$
\citep{wes09}. For a single crystal with nominally 5\% Ge substituted for Si
showing a five times larger residual resistivity, $\Delta \rho(T) = A'T$ is
more strictly observed all the way up  to $T=10$ K \citep{cus03},
which suggests a degree of disorder modulation to this `strange-metal'
behavior. If one restricts to a relatively narrow temperature range, $\Delta
\rho(T) $ has also been fit with other exponents, $A^{\prime\prime} T^{\alpha}$
($\alpha<1$); this includes the case of $\alpha=3/4$ in the temperature window
0.4~K $< T < 1$~K for the high-quality crystal exploited in Fig.\ \ref{c-rho}B,
as motivated by the theory of `critical quasiparticles' \citep{woe11} (which,
however, is inapt to explain the measured asymptotic linear-in-$T$ dependence
of the resistivity at the unconventional QCP in \yrs). Most importantly,
though, for all samples studied so far, the asymptotic ($T\rightarrow 0$)
$T$-dependence of $\rho(T)$ registered at $B = B_{\rm N}$ is found to be
linear, see also \cite{ngu21}. The linear-in-$T$ coefficient, $A'$, can be
converted to a linear-in-$T$ coefficient in the scattering rate, $1/\tau$, in
a procedure based on a Drude analysis: the latter is found to be much smaller
than what appears in a Planckian form when a `background' heavy-fermion value
is used for the effective mass \citep{ngu21,tau22}.

One of the important techniques that provides new insight into such
correlation-driven phenomena is scanning tunneling spectroscopy (STS) with its
unique ability to give local, atomically resolved information that relates to
\begin{figure*}[t]
\begin{center}
\includegraphics[width=0.84\textwidth]{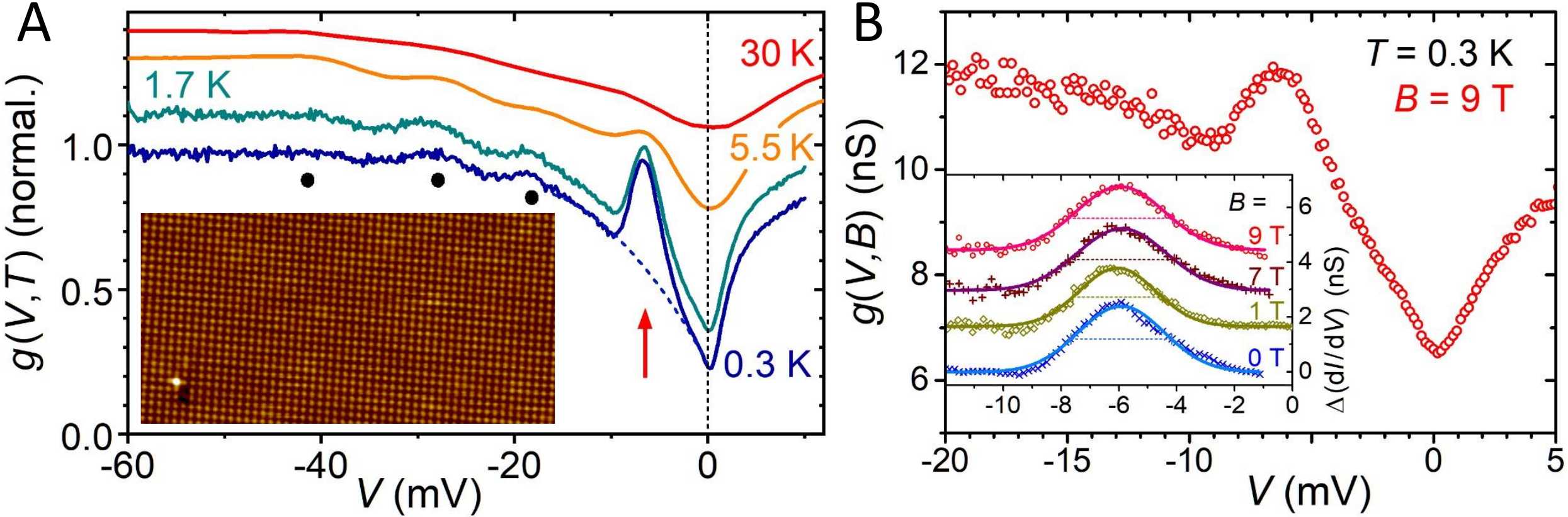}
\end{center}
\caption{Tunneling conductance $g(V,T,B)$ on Si-terminated \yrs. \textbf{(A)}
Data obtained at different temperatures and $B = 0$. $g(V,T)$-values are
normalized at $V = - 80$~mV and offset for clarity. The $-6$ meV-peak evolving
at low $T$ is marked by a red arrow; the parabola background for peak analysis
at $T =0.3$~K is shown as a dashed line. Black dots indicate features resulting
from crystalline electric field splitting of the Yb 4$f$ multiplet. Inset:
Topography visualizing the excellent surface quality (area: $20 \times 10$
nm$^2$, $V =100$ mV, $I =0.6$ nA). \textbf{(B)} Exemplary in-field data at
$T = 0.3$ K and $B = 9$ T ($\parallel c$). Inset: Data (markers) after
background subtraction with corresponding Gaussian fits (lines). A reduced
width (full width at half maximum, dashed lines) of the Gaussian at $B = 1$~T
can be recognized. All figure parts adapted from \cite{sei18}.} \label{STS}
\end{figure*}
the one-particle Green's function \citep{kir20}. However, one of the
prerequisites for successful STS measurements is the preparation and
perpetuation of clean sample surfaces. Fortunately, \yrs\ can be cleaved
nicely perpendicular to the $c$-axis in ultra-high vacuum (UHV) and at low
temperatures of about 20~K providing atomically flat surface areas of often
several hundreds of nanometers in extent. Such surfaces not only evidence the
excellent sample quality (inset to Fig.\ \ref{STS}A), they even allow to
analyze the defects to be mostly caused by Rh-atoms on Si sites \citep{wir12}.

We here focus on predominantly encountered Si-terminated surfaces. In case of
such surfaces, the Kondo-active Yb atoms are located in the fourth-to-topmost
layer which prevents a reduced screening of the Yb local moments at the surface
\citep{ale15} and ensures a predominant study of bulk properties by STS. The
latter is clearly evidenced by the observation of crystal field excitations in
the tunneling spectra \citep{ern11} at energies in excellent agreement with
inelastic neutron scattering data \citep{sto06a} (black dots in Fig.\
\ref{STS}A). Moreover, Si-terminated surfaces promote predominant tunneling
into the conduction band as compared to the 4$f$ quasiparticle states and
thereby simplify the analysis of the obtained spectra as co-tunneling can be
neglected \citep{ern11,kir20}. In this simplified picture, the successive
formation of the single-ion Kondo effect upon lowering the temperature results
primarily in a modification of the density of states of the conduction band
seen as a strong decrease of the tunneling conductance $g(V,T)$ for $V$ small
enough to not break up the quasiparticles. This process is observed to commence
at around 100~K and coincides with the onset of local Kondo screening involving
excited crystalline electric field levels as concluded from entropy estimates
\citep{cus03}. Upon cooling to below the single-ion Kondo temperature
$T_{\rm K} \simeq 25$~K the 4$f$ electrons condense into the Kramers doublet
ground state and the Kondo lattice develops. This is reflected in the STS data,
Fig.\ \ref{STS}A, by the strong development of a peak at around $-6$~meV.
Notably, the position in energy of this peak does not depend on temperature.

The relation of this $-6$~meV-peak to the Kondo \emph{lattice} is supported
by calculations: Results of a multi-level finite-$U$ non-crossing approximation
\citep{ern11} which does not consider intersite Kondo correlations captures
the temperature evolution of the zero-bias conductance dip remarkably well but
provides no indication for a peak at $-6$~meV. Conversely, renormalized band
structure calculations \citep{zwi11} which treats the fully renormalized Kondo
lattice ground state finds a partially developed hybridization gap at slightly
smaller energy in the quasiparticle density of states (which is complementary
to the here measured density of states of the conduction band within the Kondo
regime).

Albeit this Kondo lattice peak sets in at around $T_{\rm K}$, i.e. $T_{\rm coh}
\approx T_{\rm K}$ as mentioned above, it only slowly increases in height down
to about 3~K $\approx 0.1\, T_{\rm coh}$, cf. tunneling conductance at 5.5~K
in Fig.\ \ref{STS}A. This, along with a further decrease of $g(T)$ around zero
bias, may explain why single-ion descriptions can often be applied to
temperatures well below $T_{\rm K}$ despite neglected lattice Kondo effects
\citep{col85,sun13}. Only below about 3~K, the $-6$~meV-peak gains considerably
in height indicating dominant lattice Kondo correlations at these low
temperatures. This is in line with magneto- \citep{Fri10} and thermal transport
\citep{har10} investigations. In particular, the comparison of the STS data
with thermopower measurements indicates the formation of a medium-heavy Fermi
liquid down to about 3~K while strong non-Fermi liquid behavior sets in only
below this temperature. Apparently, quantum criticality only sets in if there
is sufficient buildup of lattice Kondo correlations at low enough temperatures
\citep{sei18}. A similar conclusion is suggested by resonant angle-resolved
photoemission spectroscopy on CeRhIn$_5$ \citep{che18}.

\begin{figure*}[t]
\begin{center}
\includegraphics[width=0.75\textwidth]{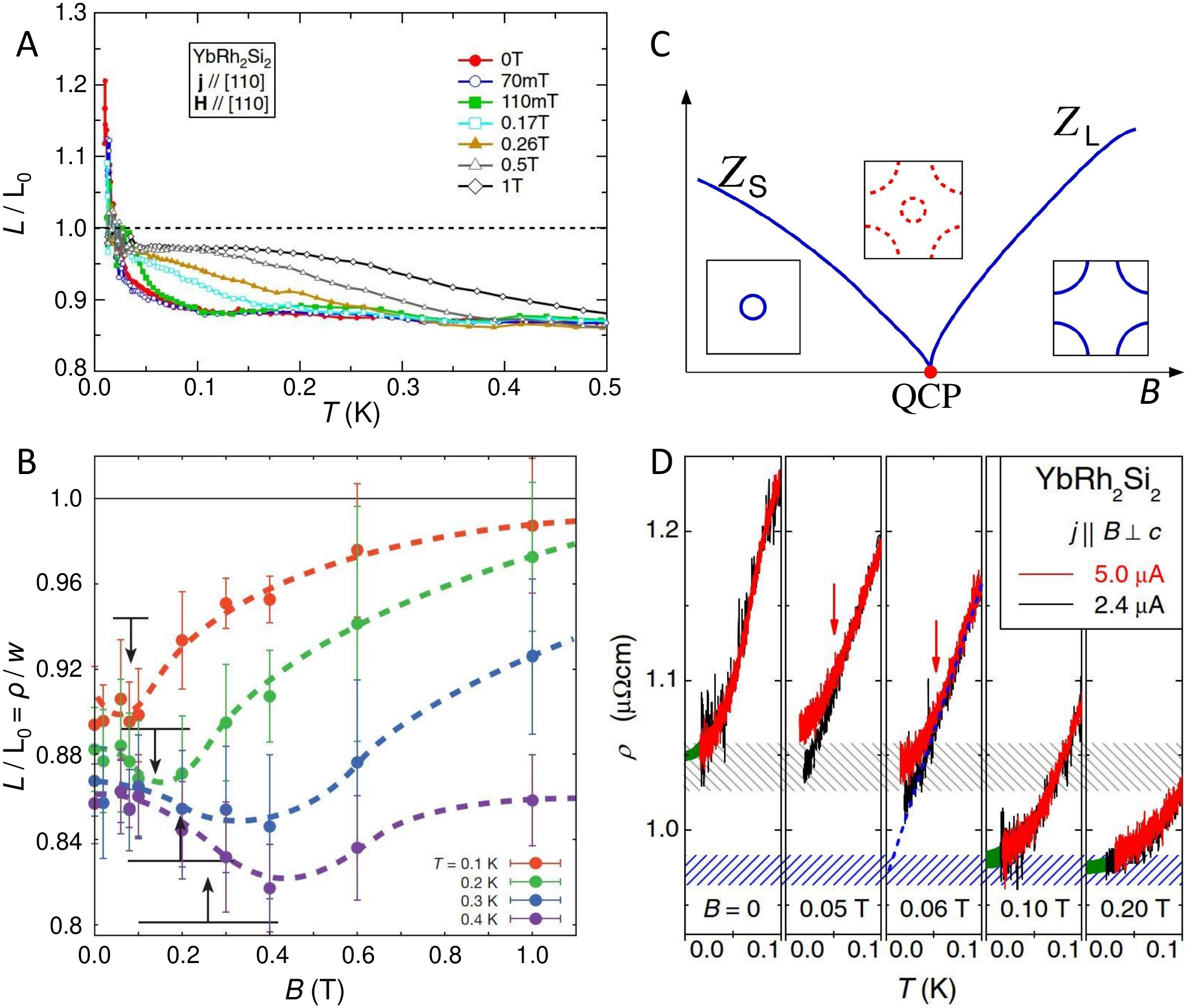}
\end{center}
\caption{\textbf{(A)} Lorenz ratio $L(T)$/L$_0$ {\it vs} $T$ for an \yrs single
crystal at $B =$ 0 and at various finite fields $B \perp c$. Reproduced
with permission from \cite{pou14}, Copyright (2014) The Physical Society
of Japan. \textbf{(B)} $L(T)$/L$_0$ {\it vs} $B$ ($\perp c$) isotherms, $T \ge
0.1$ K (paramagnetic regime, purely electronic heat transport, no bosonic
contribution to $\kappa$). Reproduced from \cite{pfau12}. \textbf{(C)}
Schematic sketch showing coexistence of small and large Fermi surfaces.
Reproduced from \cite{pfau12}. \textbf{(D)} $B_{\rm N} = 0.059$ T. Black (small
$j$): no heating. $B = 0$, $[\rho(T) - \rho_{01}] = A_1 T^2$; $B = 0.2$~T,
$[\rho(T) - \rho_{02}] = A_2 T^2$ (extrapolations to $T = 0$ shown in green)
with $\rho_{01} > \rho_{02}$ (different carrier densities, small {\it vs} large
FS). $A_1 > A_2$ (see Fig.\ \ref{c-rho}B). $B =$ 0.05 T ($\le B_{\rm N}$) resp.
$B =$ 0.06 T ($\ge B_{\rm N}$), $[\rho(T) - \rho_{01}] \approx [\rho(T) -
\rho_{02}] \approx A'T$; red (large $j$): heating on the approach of the QCP
due to additional inelastic scatterings discussed in the text. Reproduced from
\cite{steg14}.} \label{lorenz}
\end{figure*}

STS was also conducted at 0.3~K for magnetic fields applied $B \parallel c$,
see exemplary data for $B = 9$~T in Fig.\ \ref{STS}B. It should be noted that
renormalized band structure calculations \citep{zwi11} predict a smooth
quasiparticle disintegration up to well above 30 T in \yrs\ while AF order is
already suppressed at $B \approx 0.6$~T for $B \parallel c$. The position in
energy of the Kondo lattice signature peak at $-6$~meV is not influenced by
applying a magnetic field, further supporting its attribution. After background
subtraction it can be well fitted by a Gaussian, cf.\ inset of Fig.\
\ref{STS}B. The width of the Gaussian fit is somewhat reduced at $B = 1$~T
compared to peak widths at zero field and at several Tesla. Here we note that
a field magnitude of 1 T is close to the $T^*$-line (cf. Fig.\ \ref{phase})
for $T = 0.3$ K and $B \parallel c$. Therefore, the reduced peak width at
$B = 1$~T is consistent with a reduced quasiparticle weight related to quantum
criticality, see also Fig.\ \ref{lorenz}C. However, these STS data on
their own do not allow to distinguish between different scenarios for
quantum criticality and should be extended to lower temperatures.

Consequently, the question concerning the nature of the ‘local’ QCP in \yrs\
and the associated critical excitations remains. To answer this question,
combined thermal and electrical transport investigations on \yrs\ single
crystals were carried out down to 25 mK at zero field, close to $B_{\rm N}$
and up to $B = 1$~T ($\gg B_{\rm N}$) \citep{pfau12}. Subsequently, Pourret
et al. were able to extend such measurements down to even 8 mK \citep{pou14}.
The main quantity to study in this context is the Lorenz number $L = \rho
\kappa / T$, where $\rho$ is the electrical resistivity and $\kappa$ the
thermal conductivity. By defining the thermal resistivity as $w =$ L$_0 T /
\kappa$, with L$_0 = (\pi k_{\rm B})^2 / 3e^2$ being Sommerfeld’s constant, the
Lorenz ratio $L(T)/$L$_0$ can be written as $L/$L$_0 = \rho / w$. If the
Wiedemann-Franz law is valid ($L(T\rightarrow 0) /$L$_0 = 1$), which strictly
holds for elastic scattering only, the \emph{residual} electrical and thermal
resistivities turn out to be identical: $\rho_0 / w_0 = 1$.

Very different phenomena can cause a violation of this law: (i) Fermionic
excitations like spinons, i.e., charge-neutral heat carriers, may lead to
$L(T\rightarrow 0) /$L$_0 > 1$. This was indeed concluded from measurements
on, e.g., LiCuVO$_4$ \citep{par04}. (ii) Alternatively, an enhanced $w(T)$ can
lead to $L/$L$_0 < 1$, as frequently observed at \emph{finite} temperature
with dominating inelastic scatterings of the charge carriers, like the ones
from acoustic phonons. In the zero-temperature limit, however, inelastic
scatterings have to disappear. To our knowledge, before 2012 this latter
kind of violation of the Wiedemann-Franz law, $L(T \rightarrow 0)/$L$_0 < 1$,
has never been convincingly established. For example, for the
quasi-two-dimensional (2D) heavy-fermion metal CeCoIn$_5$, where an AF QCP was
suspected \citep{sin07,zau11} but not identified, $L(T)/$L$_0$ was extrapolated
to about 0.8 as $T \rightarrow 0$ for $c$-axis transport at $B \approx B_{c2}
\approx 5$ T, while it approaches $L(T)/$L$_0 \approx 1$ for in-plane
($\perp c$) transport \citep{tan07}. This result was ascribed to the action of
anisotropic spin fluctuations, although as $T \rightarrow 0$, spin fluctuations
as bosonic excitations must disappear as well. Subsequently, the observations
by Tanatar et al.\ could be consistently explained within the framework of
quasi-2D transport \citep{smi08}.

In the following, we describe the violation of the Wiedemann-Franz law in
\yrs\ with the aid of Fig.\ \ref{lorenz}, see also \citep{smi18}. As shown in
Fig.\ \ref{lorenz}A at $B = 0$ and 0.07 T, $L(T)/$L$_0 \approx 0.87$ in an
extended temperature window (0.1 K $< T <$ 0.5 K) \citep{pou14}. In this
$T$-range, the underlying electrical and thermal resistivities depend linearly
on $T$, so that the electronic Lorenz ratio $L_{\rm el} /$ L$_0$ is
temperature-independent. While for $\rho(T)$ the ‘strange-metal' behavior
persists to the lowest accessible temperature, and most likely to absolute
zero, an additional bosonic contribution $\kappa_{\rm m}(T)$ (by magnons at
$B = 0$, resp. paramagnons at $B =$ 0.07 T) is added to the electronic thermal
conductivity $\kappa_{\rm el}(T)$ at $T \le 0.1$ K which means that here, the
total thermal resistivity $w(T)=[\kappa_{\rm el}(T) + \kappa_{\rm m}(T)]^{- 1}$
drops, and a distinct upturn develops in $L(T)/$L$_0$, as clearly seen in Fig.\
\ref{lorenz}A. Because of its bosonic nature, this additional term has to
vanish as $T \rightarrow 0$, whereby it must pass over a maximum below the
low-$T$ limit of the experiments (8~mK). The (constant) low-$T$ value of
$L_{\rm el} /$ L$_0 \approx 0.87$, displayed in Fig.\ \ref{lorenz}A over an
extended temperature window, is also derived from the minimum value of the
$L(B)/$L$_0$ isotherm for $T = 0.1$ K, the lowest temperature at which no
interfering paramagnon contribution to $\kappa (T)$ exists (red data points in
Fig.\ \ref{lorenz}B). These data were obtained with a different set up on a
different single crystal. We thus conclude that the ratio $\rho_0 / w_0 =
\rho_0 / w_{\rm el,0}$ is reduced by about 10\% compared with unity, the
value expected from the Wiedemann-Franz law. Fig.\ \ref{lorenz}B demonstrates
that below 1 K, the Lorenz ratio is generally less than unity which
implies predominating inelastic scattering processes, i.e., the ordinary
small-angle electron-electron and electron-spin fluctuation scatterings. As
already mentioned, in this low-temperature range, a broad minimum shows up in
the $L(B)/$L$_0$ isotherms displayed in the figure, which points to \emph{an
additional inelastic scattering process}. This minimum is found to occur around
the $T^*(B)$-line and to become narrower upon cooling. We therefore consider it
to represent the dynamical origin of local quantum criticality in \yrs\ by
ascribing it to scatterings that are associated with the transformation
between a small and a large Fermi surface, which coexist on either side of
$T^*(B)$ all the way down to the QCP ($T = 0, B = B_{\rm N}$). As displayed in
Fig.\ \ref{lorenz}C, the quasiparticle weights on both sides are smoothly
vanishing as $T \rightarrow 0$, whereby the minimum in $L(B)/$L$_0$ becomes a
delta function, resulting from \emph{fermionic} quantum critical fluctuations
(which is a rare case, as in most scenarios quantum critical fluctuations are
of \emph{bosonic} origin). Apparently, these critical fluctuations are
instrumental to enhance the residual thermal resistivity by about 10\% over
its electrical counterpart.

Several groups have reported very similar experimental data compared with
those by \cite{pfau12}, but questioned the interpretation sketched above. The
key problem is the correct treatment of the bosonic term $\kappa_{\rm m}(T)$.
This term was just ignored by \cite{mac13,rei14}, i.e., here the measured
thermal conductivity was erroneously regarded as the electronic contribution
in the whole low-temperature range of the experiments, down to 40 mK. On the
other hand, \cite{tau15}, who gave a detailed interpretation of the data
previously published by \cite{pou14}, consider $\kappa_{\rm m}(T)$ to set in
at a temperature as low as 30 mK, although the data \citep{pou14} clearly
prove this to occur already at about 0.1 K (see Fig.\ \ref{lorenz}A).
Therefore, on extrapolating the data to $T = 0$ from just above 30 mK where
$\kappa_{\rm m}(T)$ \emph{dominates}, they miss the intrinsic value
$L_{\rm el}(T \rightarrow 0)/{\rm L}_0 \approx 0.9$ and instead obtain
accidently $\approx 0.97$. This leads to their false claim that the
Wiedemann-Franz law holds in \yrs. A violation of the Wiedemann-Franz law was
subsequently also reported for another heavy-fermion metal, YbAgGe
\citep{dong13}.

In Fig.\ \ref{lorenz}D, the conclusions drawn from the heat-conduction study
discussed above are nicely confirmed by results of measurements of the
electrical resistivity \citep{lau13}. Close to the critical field
$B_{\rm N} = 59$ mT and at sufficiently low temperatures, the sample under
investigation becomes heated by a moderate current due to its deteriorated
heat conductivity. No heating is observed when applying a low enough current.
Away from $B_{\rm N}$, in the AF phase at $B = 0$ as well as in the
paramagnetic phase at 0.1 and 0.2 T, the resistivity follows the Fermi
liquid-type $T^2$-dependence, independent of the size of the here
investigated currents. Upon approaching $B_{\rm N}$ from either side, at the
lowest accessible temperatures ‘strange-metal' behavior, characteristic of the
local QCP, is observed at low current. If these linear $T$-dependences of
$\rho(T)$ obtained at $2.4\, \mu$A are extrapolated to $T = 0$, $\rho(T)$ ends
up at very different values of the residual resistivity. In particular,
\begin{figure*}[t]
\begin{center}
\includegraphics[width=0.72\textwidth]{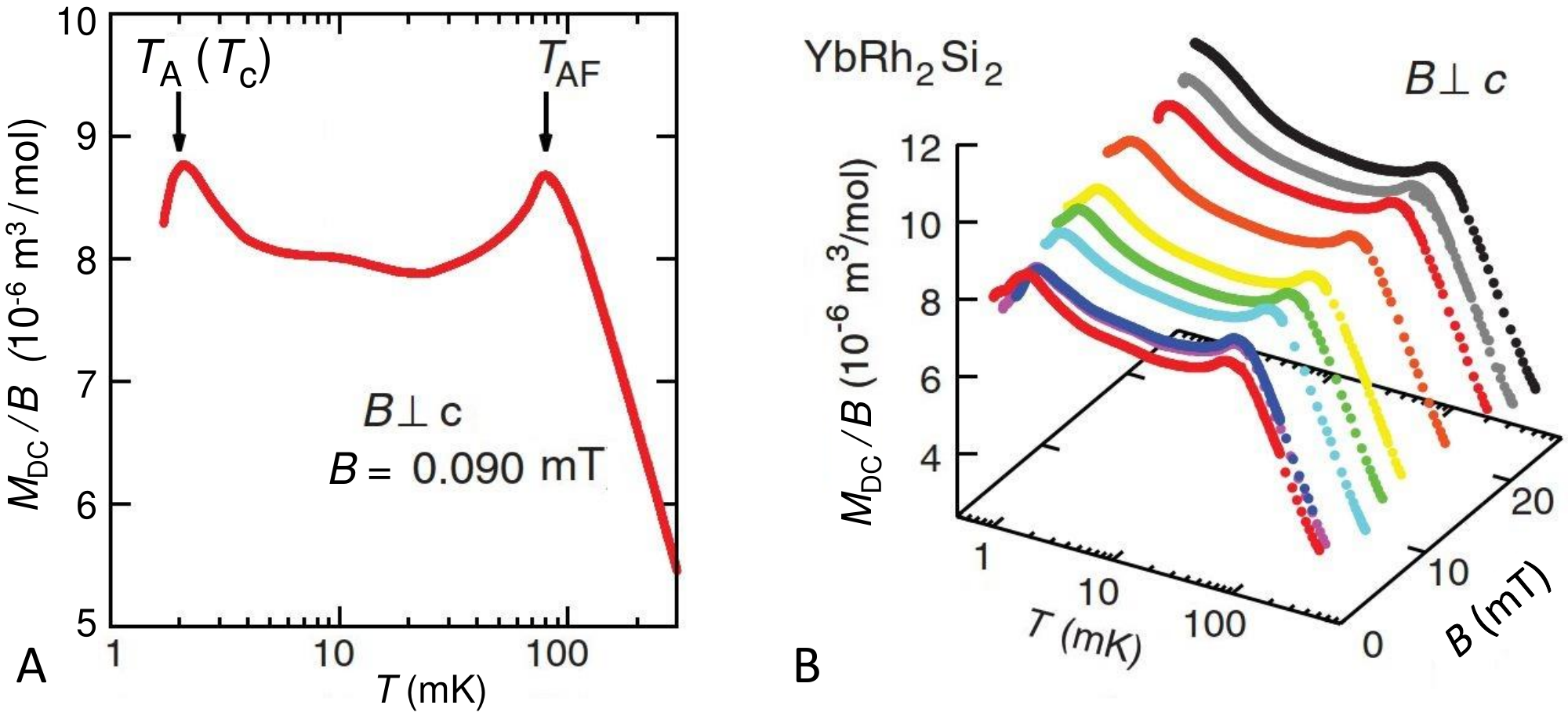}
\end{center}
\caption{Field-cooled (fc) DC magnetization $M_{\rm DC}$, normalized at 1.1 mT,
in dependence on temperature $T$ for magnetic field $B \perp c$. After cooling
the sample at the magnetic measuring field $B$ to below 1~mK, the magnetization
curve was taken on the warmup of the nuclear stage. \textbf{(A)} $B =$ 0.090
mT. \textbf{(B)} $B \le 23$ mT. Reproduced from \cite{sch16}.} \label{MdcTB}
\end{figure*}
the so extrapolated $\rho_0$ values match nicely with those obtained by
extrapolating the Fermi liquid-type $T^2$-dependences of $\rho(T)$ found at
$B = 0$ and way above $B_{\rm N}$, respectively. We consider this jump in
$\rho_0$  as a direct visualization of the abrupt change in the charge-carrier
density of \yrs\ on field tuning through the Kondo-destroying QCP.

\section{Competition between Nuclear and Primary
4\lowercase{\textit{f}}-Electronic Order, Emergence of Hybrid A-Phase and
Superconductivity}
\label{sec2}
When \yrs\ single crystals were investigated by resistivity measurements at
$T > 10$ mK, and specific-heat as well as susceptibility measurements at $T >$
18~mK, no superconductivity could be detected \citep{cus03}. The most natural
explanation for this is that superconductivity becomes suppressed by the AF
order which forms at $T_{\rm N} =$ 70 mK. To find out whether superconductivity
in \yrs\ shows up at $T \le 10$~mK, magnetization, susceptibility and
specific-heat measurements have been carried out down to temperatures as low
as 0.8 mK by using a nuclear demagnetization cryostat providing a base
temperature of 400 $\mu$K \citep{sch16}. Here, a total of 5 different
single crystals was investigated. In Fig.\ \ref{MdcTB}A, the temperature
dependence of the field-cooled (fc) DC-magnetization, measured at a magnetic
field of 0.09 mT, reveals two phase-transition anomalies at $T_{\rm N} \approx
70$ mK and $T_{\rm A}$ ($T_{\rm c}$) $\approx 2$ mK. While the peak at 70 mK
illustrates the AF 4$f$-electronic transition, the one at 2 mK marks the
transitions into both nuclear-dominated hybrid AF order (‘A phase') and
heavy-fermion superconductivity, as discussed below. A blow-up of the data
near the 2 mK-peak at fields below 4 mT indicates that the onset of hybrid
order precedes that of superconductivity, with $T_{\rm A} - T_{\rm c}$ being
less than $0.1 T_{\rm A}$ \citep{sch16}. Fig.\ \ref{MdcTB}B illustrates how
these phase-transition anomalies evolve with increasing magnetic field. The
position of the low-temperature anomaly becomes gradually reduced until the
latter cannot be resolved anymore above 23 mT, whereas $T_{\rm N}$ is robust
in this whole field range. Clearly resolved is an increase in field-cooled
magnetization, fc-$M_{\rm DC}(T)$, upon cooling to below about 20 mK, which
indicates a weakening of the staggered magnetization in the primary AF phase.
At $T \approx 10$ mK, a significant decrease in the absolute slope of
fc-$M_{\rm DC}(T)$ is observed.

\begin{figure*}[t]
\begin{center}
\includegraphics[width=12cm]{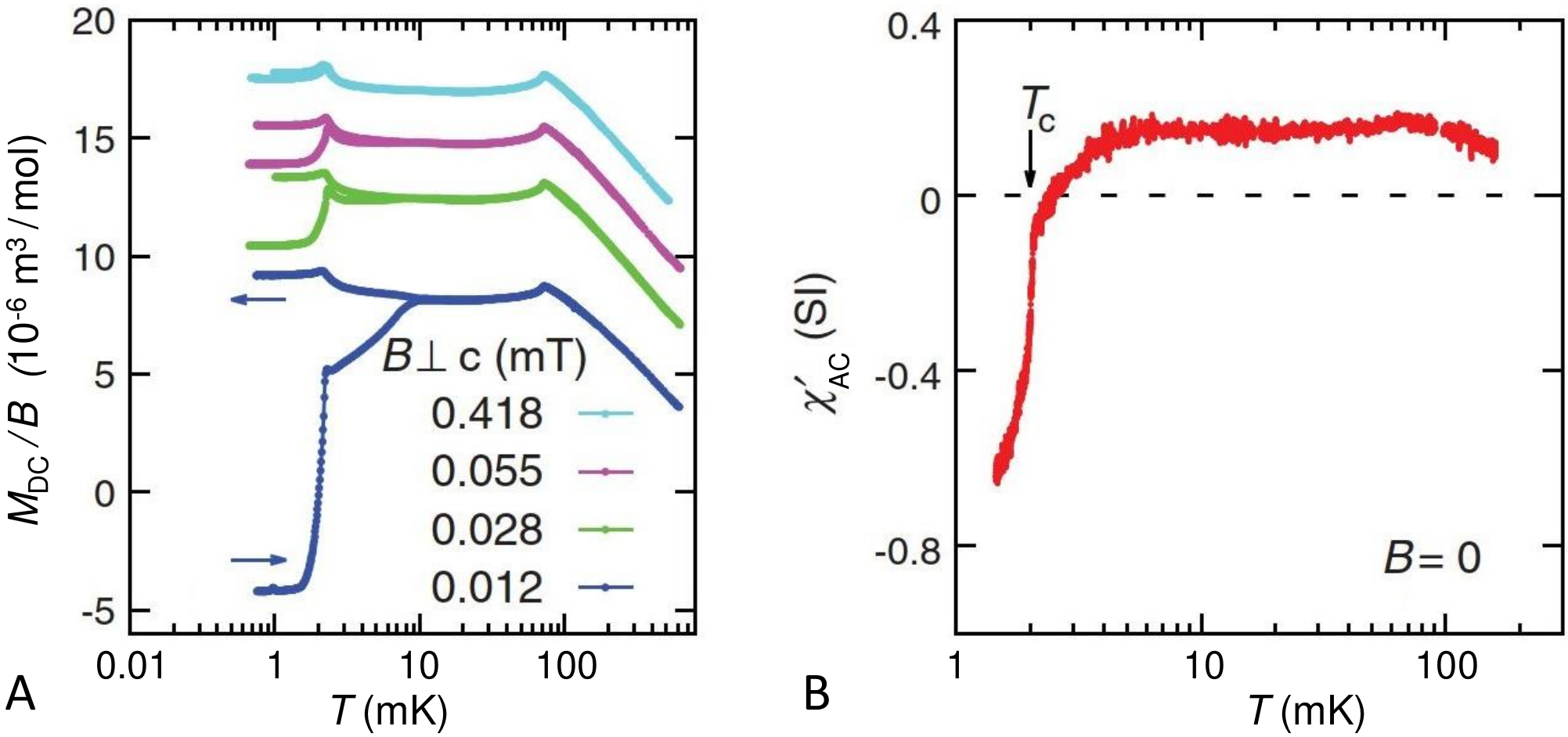}
\end{center}
\caption{\textbf{(A)} fc- and zfc-$M_{\rm DC}(T)$ curves, normalized to
1.1~mT and offset for the sake of clarity, between 0.012 and 0.418 mT. The
data were taken in a special measuring cell which allowed for the compensation
of the earth field to obtain a smallest magnetic measuring field of 0.012~mT.
Note that the vertical scale is larger by more than a factor of 5 compared to
that in Fig.\ \ref{MdcTB}A. \textbf{(B)} Real part of the AC susceptibility
$\chi_{\rm AC}^{\prime}(T)$ at $B \approx 0$. Reproduced from \cite{sch16}.}
\label{M-ac}
\end{figure*}

Figure \ref{M-ac}A displays the zero-field-cooled (zfc-) and fc-$M_{\rm DC}(T)$
curves taken up to $B = 0.418$ mT in a specially shielded setup (different
from the one used to obtain the data of Fig.\ \ref{MdcTB}). The data
registered at the lowest field, $B =$ 0.012 mT, illustrate how the experiment
was performed: one starts at $T > 10$ mK by cooling the sample in zero field
to the lowest temperature, $T = 0.8$ mK. Then, the field is applied and the
zfc curve is recorded on warming to above 10 mK. Cooling again with field
applied yields the fc curve. The zfc-$M_{\rm DC}(T)$ curve, which
separates abruptly from the fc curve at $T \approx 10$ mK, indicates a
shielding signal which is increasing almost linearly upon cooling and assumes
a value of not more than 20\% just above $T_{\rm c} = 2$ mK. This is followed
by a sharp, pronounced drop at $T_{\rm c}$ and a robust diamagnetic response
at $T \le 1$ mK. These data are well reproduced by the results of the AC
susceptibility obtained under nearly zero-field conditions, Fig.\
\ref{M-ac}B. In the $\chi_{\rm AC}(T)$ data partial shielding below $T \approx
10$ mK and the AF phase transition at $T_{\rm N} \approx 70$ mK are resolved
as well.

We now turn to the peak in the fc-$M_{\rm DC}(T)$ curve at 2 mK which reveals
a pronounced decline $\Delta M$ of about 0.075~$\mu_{\rm B}$ per Yb down to
0.8~mK (Fig\ \ref{MdcTB}A). As will be discussed below, about 25\% of this
decline on the low-$T$ side of the 2~mK-peak should be attributed to the onset
of the nuclear-4$f$ electronic hybrid ‘A-phase', leaving about 75\% of
this being due to the Meissner effect. This corresponds to a Meissner volume
amounting to only about 2\% of the full shielding signal which seems to be
quite a small value; however, owing to vortex pinning this is typical for bulk
type-II superconductors. After destroying the pinning centers by powdering and
subsequent annealing, the sample should exhibit a substantially increased
Meissner volume when measured below the lower critical field $B_{c1}$, see,
e.g., \cite{ste79,rau82}. As also inferred from Fig. \ref{M-ac}A, the shielding
signal in zfc-$M_{\rm DC}(T)$ has become extremely weak at a field as low as
0.418~mT, cf.\ the discussion below. By contrast, the jump in
fc-$M_{\rm DC}(T)$, $\Delta M$, is robust, hinting at the existence of
\emph{bulk} superconductivity with $T_{\rm c} \ge 0.8$ mK up to $B \leq 23$~mT,
see Fig.\ \ref{MdcTB}B.

\begin{figure}[b]
\begin{center}
\includegraphics[width=0.48\textwidth]{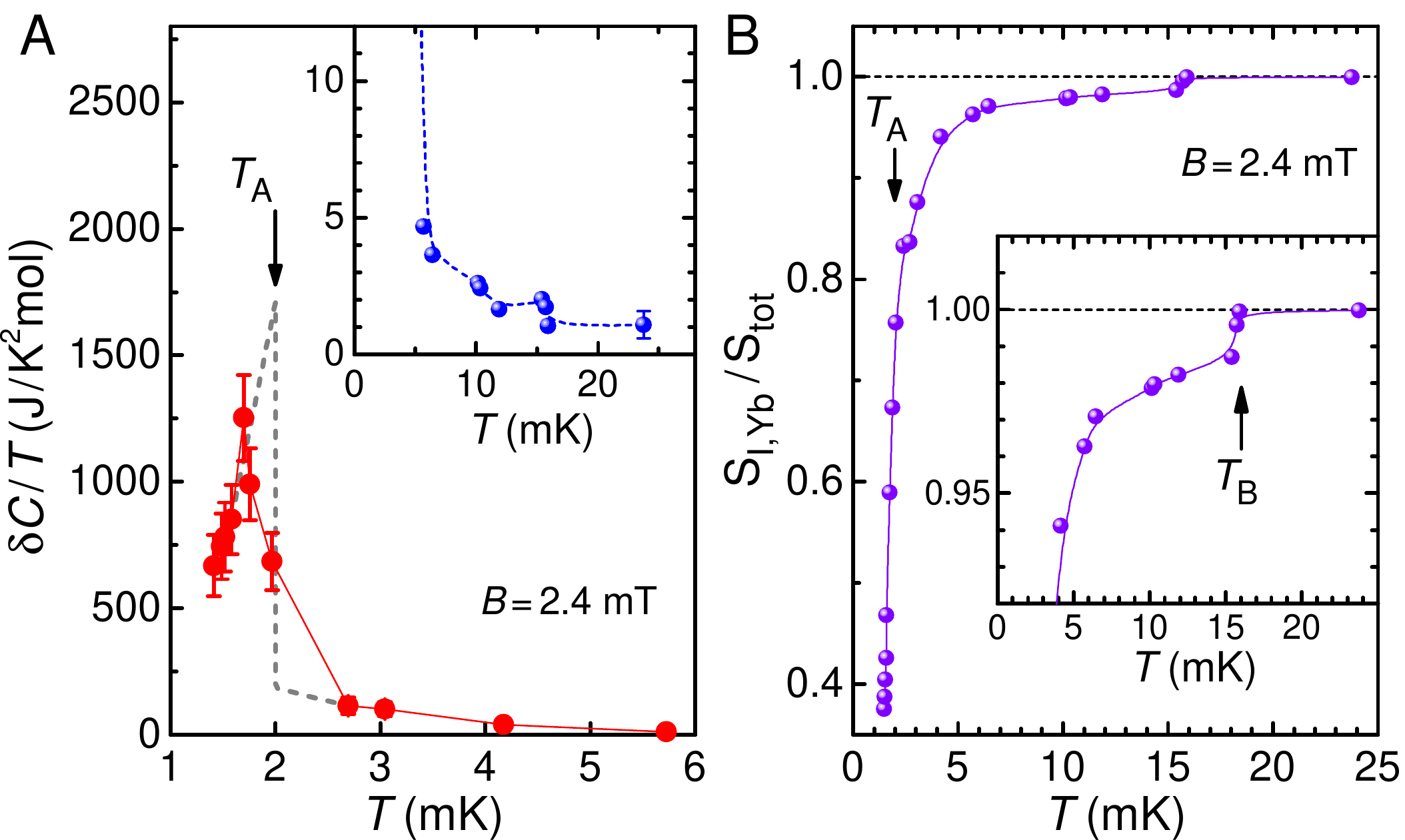}
\end{center}
\caption{\textbf{(A)} Molar spin specific heat of \yrs\ in an external
magnetic field $B = 2.4$~mT plotted as $\delta C/T$ vs $T$ at $T < 6$~mK.
The data were obtained by the common heat-pulse as well as the relaxation
methods with the fc-$M_{\rm DC}(T)$ dependence of \yrs\ used as an internal
thermometer. In the latter case, the heat capacity $\tilde{C}$ (including
addenda contributions) could be determined by the relaxation time $\tau =
\tilde{C}\, R_{\rm th}$, with $R_{\rm th}$ being the thermal resistance of the
`weak link' (between sample and nuclear stage). For details, see \cite{sch16}.
The inset shows the same quantity on a smaller scale for $T < 30$~mK.
\textbf{(B)} Yb-derived molar nuclear spin entropy $S_{\rm I,Yb}/ S_{\rm tot}$
vs.\ $T$ in units of $S_{\rm tot}$ where $S_{\rm tot}$ is its value at
sufficiently high $T$. Inset: Zoom into the same data at $T \ge 4$~mK to
emphasize the jump at $T_{\rm B}$.}  \label{SpecHeat}
\end{figure}

The coefficient of the molar spin specific heat, $\delta C(T)/T$, obtained
after subtracting a huge nuclear quadrupolar contribution (for $B = 0$) from
the raw data taken at 2.4 mT, is shown below 6 mK in Fig.\ \ref{SpecHeat}A.
$\delta C(T)$ mainly consists of the contributions by the Yb-derived nuclear
spins ($S = 1/2$ for $^{171}$Yb ions with a natural abundance of 14.3\% as
well as $S = 5/2$ for $^{173}$Yb ions with 16.1\% abundance). Note that
neither the $^{100}$Rh nor the $^{29}$Si nuclear spins contribute to the
specific heat above $T = 1$~mK, because they assume their full Zeeman entropies
already below this temperature. In addition to the nuclear spin contributions,
there is a small one by the 4$f$-electronic spins, $C_{4f}(T)$. Since the
effect of a magnetic field on the nuclear quadrupole contribution is only of
higher order, one can use these $\delta C(T)/T$ data, subtracted by
$C_{4f}(T)/T$, to estimate the molar Yb-derived nuclear spin entropy
$S_{\rm I,Yb}(T)$ (for $B =$ 2.4 mT). Clearly seen in Fig.\ \ref{SpecHeat}A
is a huge, broadened phase-transition anomaly of mean-field type. The latter
can be replaced, under conservation of entropy, by a jump which yields a phase
transition temperature of $T_{\rm A} = 2$~mK (at $B = 2.4$~mT) and a jump
height $\Delta C / T_{\rm A}$ of $\approx 1700$ J/K$^2$mol. This exceeds
$\Delta C / T_{\rm c}$ observed at the transition temperature of typical
heavy-fermion superconductors by more than a factor of 1000 and indicates that
the A-phase transition is predominantly due to nuclear degrees of freedom.
Since an additional measurement at 59.6 mT revealed $T_{\rm A}$ to be shifted
to below the lowest accessible temperature of 0.8~mK, this $T_{\rm A}$-anomaly
marks the transition into a state of antiferromagnetically ordered nuclear
spins. In the inset of Fig.\ \ref{SpecHeat}A, $\delta C(T)/T$ is displayed on
a largely expanded vertical scale between 6 -- 23 mK, which now contains
additional (in comparison to \cite{sch16}) data for $T > 12$~mK. At $T \ge
18$~mK, where all nuclear spin components are negligible, these data agree
well with previous results for $C_{4f}(T)/T$ \citep{cus03}, cf.\ Fig.\
\ref{c-rho}A. The anomaly visible at $T_{\rm B} \approx 16$~mK can also be
recognized in Fig.\ \ref{SpecHeat}B, where the temperature dependence of the
entropy of the Yb-derived nuclear spins, $S_{\rm I,Yb}(T)$, is displayed in
units of its total (high-$T$) value, $S_{\rm tot}$. This anomaly shall be
discussed in more detail in the following section.

\begin{figure}[t]
\begin{center}
\includegraphics[width=8cm]{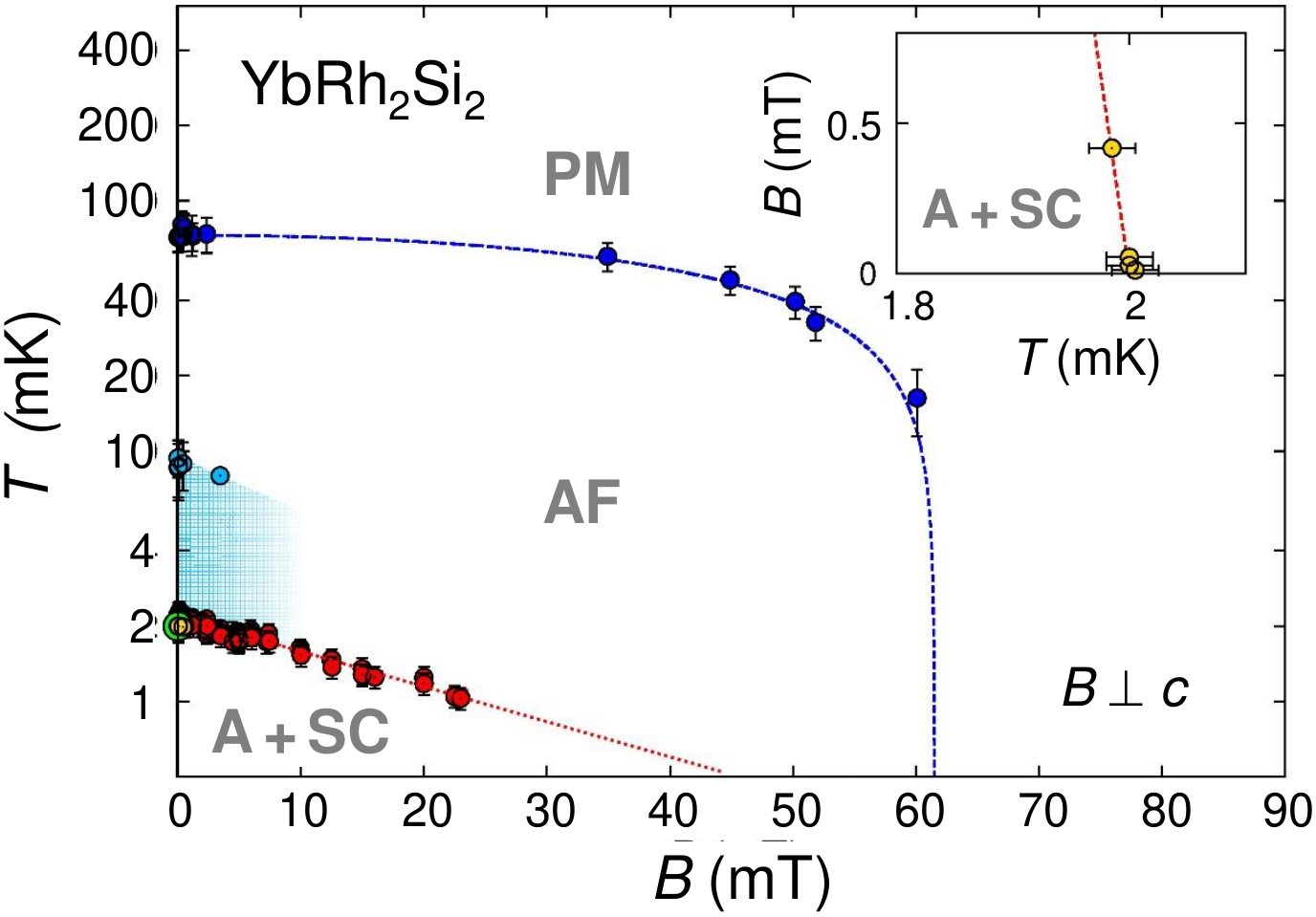}
\end{center}
\caption{Phase diagram of \yrs. Dark blue dots and line: boundary of the
primary  AF order. Light blue dots and shaded area: partial superconducting
shielding in previously labelled `B-phase', i.e., in insulated droplets of
A-phase which form below $T_{\rm B} \approx 16$~mK, see text. Red data points:
Position of low-$T$ peak in fc-$M_{\rm DC}(B)$, see Fig.\ \ref{MdcTB}. Green
circle: superconducting $T_{\rm c}$ at $B = 0$ from $\chi_{\rm AC}(T)$, see
Fig.\ \ref{M-ac}B. Yellow circles: superconducting $T_{\rm c}$ from
zfc-$M_{\rm DC}(T)$, Fig.\ \ref{M-ac}A, blown up in the inset to illustrate
the huge absolute value of the initial slope of $B_{\rm c2}(T)$ at
$T_{\rm c}$, $|B_{\rm c2}'| \approx 25$~T/K. Figure adapted from \cite{sch16}.}
\label{lowTphase}
\end{figure}

The temperature-magnetic field phase diagram in Fig.\ \ref{lowTphase}
indicates the various low-$T$, low-$B$ phases of \yrs, i.e., the primary
AF phase (blue dots and dashed line), the so-called ‘B-phase' (light blue
shading), the A-phase and superconductivity
whose transition temperatures are closely spaced and jointly displayed by
the red dots, designating the positions of the low-temperature peaks in
fc-$M_{\rm DC}(T)$, see Fig.\ \ref{MdcTB}. The green symbol at
$B = 0$ represents the superconducting phase transition observed in
$\chi_{\rm AC}(T)$, and the yellow ones, partly hidden by the former, denote
the positions of the pronounced shielding signals registered at very low
fields by zfc-$M_{\rm DC}(T)$ (Fig.\ \ref{M-ac}A). These latter transition
temperatures are plotted as a function of field in the inset, yielding the
absolute initial slope of the upper critical field curve at $T = T_{\rm c}$,
$|d B_{\rm c2}(T)/ d T| = |B_{\rm c2}'| \approx 25$~T/K. An identical value
was obtained from the fc-$M_{\rm DC}(T)$ data at very low fields \citep{sch16}.
The large magnitude of $|B_{\rm c2}'|$ is typical for heavy-fermion
superconductors, based upon the ordinary 4$f$-electronic Kondo effect
\citep{ass84}. If the giant anomaly in $\delta C(T)/T$ displayed in Fig.\
\ref{SpecHeat}A manifested rather a superconducting than a nuclear-ordering
transition, one would deal with super-heavy, almost localized, quasiparticles
which were originating in a `nuclear Kondo effect' \citep{col15b}. In this
case, the spins of the conduction electrons would rather screen the Yb-derived
nuclear than the 4$f$-electronic spins. This scenario cannot be at play as it
would result in an almost infinite slope $|B_{\rm c2}'|$.

One can derive the effective $g$-factor of the A-phase, $g_{\rm eff} \simeq$
0.051, from the ratio of the transition temperature $T_{\rm A}(B=0) = 2.3$~mK
and the critical field $B_{\rm A}(T \rightarrow 0) \simeq 45$~mT (Fig.\
\ref{lowTphase}). Using the in-plane 4$f$-$g$-factor of \yrs, $g_{4f} = 3.5$
\citep{sic03}, one finds the A-phase to represent hybrid AF order comprising
a dominant ($\simeq 98.5$\%) nuclear component and a tiny 4$f$-electronic
component of $\simeq 1.5$\%, with a staggered moment of $m_J \simeq
0.018\,\mu_{\rm B}$. This is obtained with the aid of the ($T \rightarrow 0$)
saturation moment, $1.24\,\mu_{\rm B}$, as estimated from a Curie-Weiss fit
\begin{figure}[t]
\begin{center}
\includegraphics[width=8.0cm]{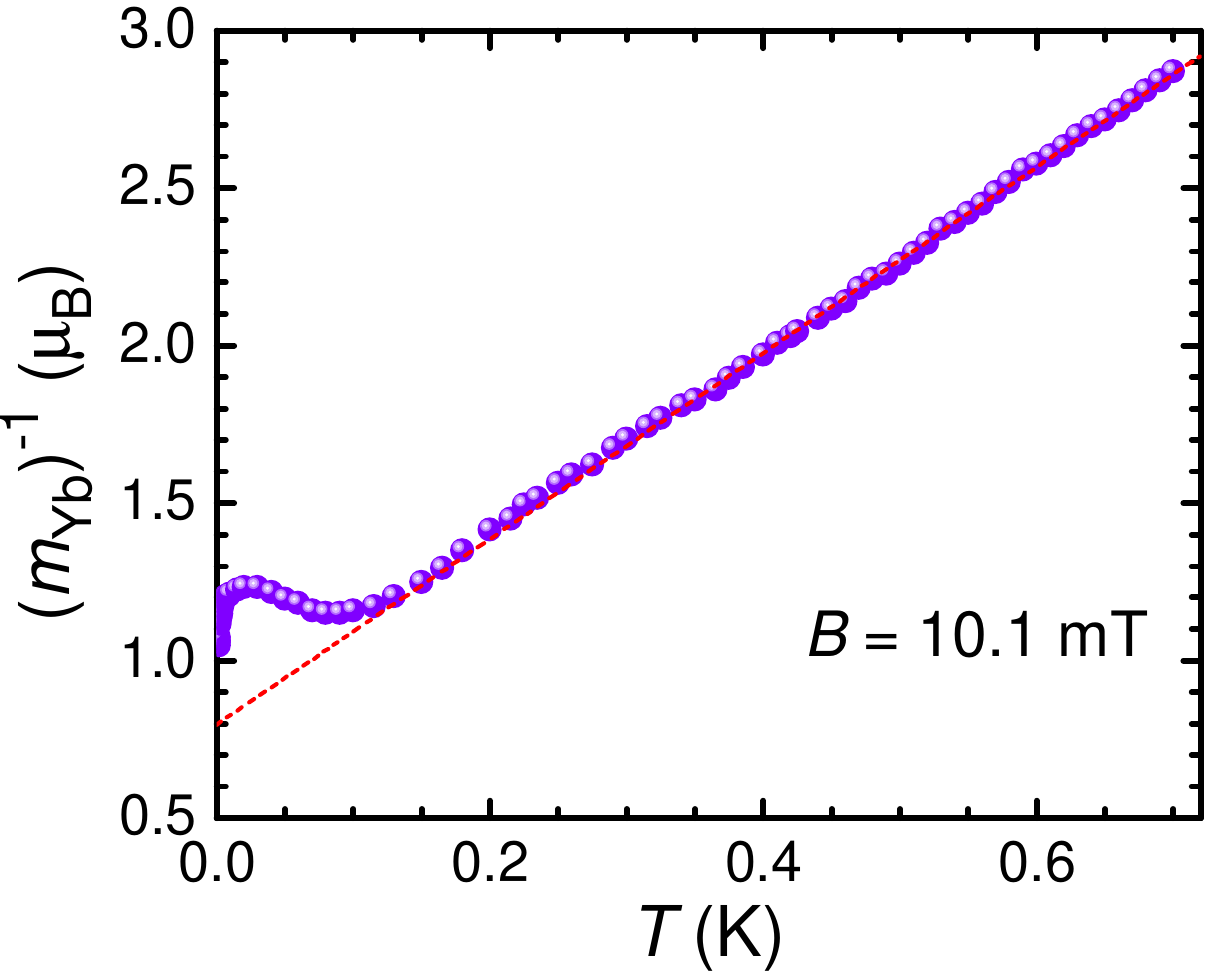}
\end{center}
\caption{Curie Weiss fit, $\mu_{\rm B} / m_{\rm Yb}$ vs. temperature at $B =$
10.1 mT. For $T \rightarrow 0$ a saturation moment $m_{\rm sat} =$ 1.24
$\mu_{\rm B}$ is estimated. The Weiss temperature $\theta \approx -0.27$~K
agrees well with the result reported by \cite{cus03} and illustrates dominant
AF correlations at low temperature \citep{ham19}.}
\label{curie}
\end{figure}
to the fc-magnetization data taken in the paramagnetic regime at an external
field of 10.1 mT, see Fig.\ \ref{curie}. The staggered moment $m_J$ exceeds
significantly that of the primary AF phase, $m_{\rm AF} \approx 0.002 \,
\mu_{\rm B}$ \citep{ish03} which may explain the `re-entrant AF order' at very
low temperatures as reported by \cite{sau18} based on measurements of Nyquist
noise. We note that a pure nuclear phase transition would not be resolved in
our magnetization measurements because of the very small nuclear moment.
Therefore, one can state that the 2 mK-peak in fc-$M_{\rm DC}(T)$ originates
in the tiny 4$f$-component of the hybrid A-phase (and additionally, to a larger
part in the Meissner signal of the superconducting transition), while the huge
anomaly in the specific-heat coefficient shown in Fig.\ \ref{SpecHeat}A is only
due to the Yb-derived nuclear spin states. We infer from the red data points
\begin{figure*}[t]
\begin{center}
\includegraphics[width=14.0cm]{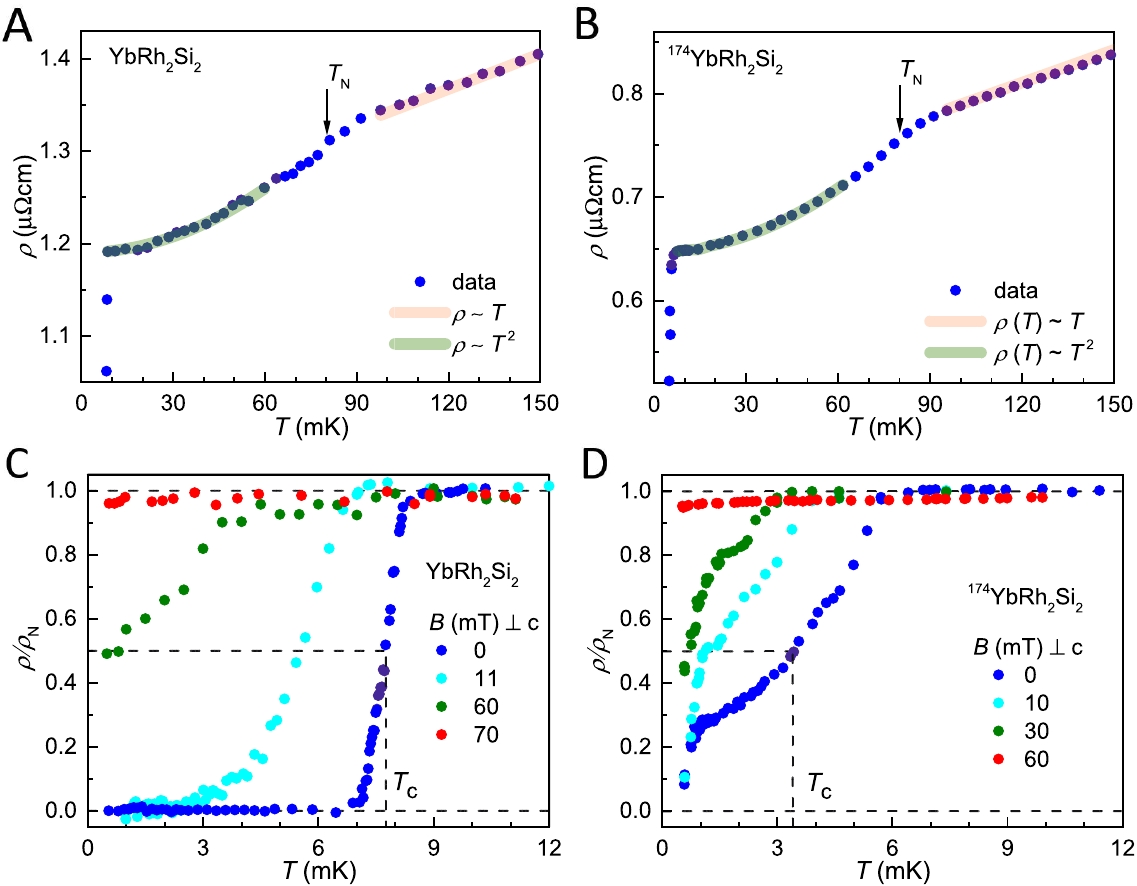}
\end{center}
\caption{Electrical resistivity as a function of temperature for \textbf{(A)},
\textbf{(C)} \yrs\ and \textbf{(B)}, \textbf{(D)} $^{174}$\yrs\ at $B = 0$
and various applied fields. Reproduced from \cite{ngu21} under a Creative
Commons Attribution 4.0 International License. Copyright 2021, The Authors,
published by Springer Nature.} \label{rhoN}
\end{figure*}
shown in Fig.\ \ref{lowTphase} and the results of Fig.\ \ref{MdcTB}B that the
existence range for bulk superconductivity is most likely extended to fields
in excess of 23~mT where the values of $T_{\rm c}$ are below 0.8~mK.

Measurements of the electrical resistivity at ultra-low temperatures are
extremely difficult in view of the unavoidable contact resistances which may
easily heat up the sample. Recently, Nguyen et al.\ were successful in
performing such measurements on \yrs\ single crystals down to 0.5 mK
\citep{ngu21}. Figure \ref{rhoN} reproduces their results on a pristine sample
of \yrs\ (A, C) as well as on a single crystal with nominally 100 at\%
of $^{174}$Yb ions which do not carry nuclear spins (denoted as $^{174}$\yrs\
in B, D). This latter sample had been prepared and intensively studied by
\cite{kne06}. For both single crystals the onset of the primary AF order at
$T_{\rm N} = 70$~mK is resolved, see Figs.\ \ref{rhoN}A and B, as is the
superconducting transition with a (mid-point) $T_{\rm c} = 7.9$~mK, following
an onset at $\simeq 9$~mK, for the \yrs\ crystal and an onset at about 5~mK
for the $^{174}$\yrs\ crystal (Figs.\ \ref{rhoN}C and D). Within the
resolution of the experiments, the resistivity reaches $\rho = 0$ at about
6.5~mK for the former sample (Fig.\ \ref{rhoN}C). In case of the nominally
nuclear-spin free $^{174}$\yrs\ sample, the onset of superconductivity at 5~mK
is followed by a broadened 75\% decrease of $\rho(T)$, which starts to decline
in a substantially steeper fashion at about 1~mK, without reaching zero at the
low-$T$ limit of 0.5~mK, suggesting $\rho = 0$ at around 0.3 -- 0.4 mK (Fig.\
\ref{rhoN}D). Under external magnetic field the transitions are shifted to
lower temperatures. As shown in Fig.\ \ref{rhoN}C, for the \yrs\ sample, the
transition is broadened at $B = 11$~mT, and zero resistivity is achieved at $T
\approx 2$~mK. Even at $B = 60$~mT a residual, very broad and incomplete
transition is recognized. The initial $T_{\rm c}(B)$ dependence is found
to be relatively steep, implying an absolute initial slope of $|B_{c2}'| \simeq
4.4$~T/K. The field dependence of $T_{\rm c}$ becomes flatter above $B \approx
11$~mT. The finite-field data for the $^{174}$\yrs\ sample displayed in Fig.\
\ref{rhoN}D show a field-induced reduction of the temperature range of the
broad initial decline in $\rho(T)$. In addition, the temperature of
0.3 -- 0.4~mK at which $\rho(T)$ is likely to vanish, appears to be almost
independent of field up to $B = 30$~mT.

To put these new results in perspective, we wish to note that
resistive transitions (at $T_{c,\rho}$) commonly probe percolative
superconductivity, with $\rho(T)$ reaching zero when the first percolating
path through the sample is formed. On the other hand, both $\chi_{\rm AC}(T)$
(with $T_{c,\chi}$) and zfc-$M_{\rm DC}(T)$ provide shielding signals due to
`networks of screening currents' in the sample surface. The thermodynamic
(bulk-) $T_{\rm c}$ is obtained through transitions in the specific heat
and/or the Meissner effect as measured by fc-$M_{\rm DC}(T)$ (with $T_{c,M}$).
For inhomogeneous superconductors one frequently finds $T_{c,\rho} >
T_{c,\chi} > T_{c,M}$. In case of the Ce-based 115-superconductors with
anisotropic transport owing to the delicate interplay of competition and
coexistence between superconductivity and AF order, an exotic type of
percolative (`textured') superconductivity has been reported \citep{par12}.
For CeIrIn$_5$, the thermodynamic $T_{\rm c}$ probed via specific heat is
0.4 K, while $T_{c,\rho} = 1.2$~K \citep{pet01}.

In the \yrs\ sample, the onset of superconductivity at very low fields is
observed at almost the same $T_{c,\rho} \approx 8$~mK in measurements of both
the electrical resistivity \citep{ngu21} and Nyquist noise  \citep{sau18}.
In this exceptional case $T_{c,\rho}$, and even the onset temperature of the
resistive transition, is \emph{smaller} than $T_{c,\chi} \approx 10$~mK.
The resistivity study yields an absolute initial slope of $B_{c2}(T)$ at
$T_{\rm c}$, $|B_{c2}'| \approx 4.4$~T/K discussed above, which is much smaller
than the value 25~T/K based on zfc- and fc-$M_{\rm DC}(T)$ results, cf.\ the
preceding Section \ref{sec2}. In addition, the thermodynamic superconducting
transition displayed by the low-temperature peak in fc-$M_{\rm DC}(T)$ is
corroborated by abrupt and large shielding signals in both zfc-$M_{\rm DC}(T)$
and $\chi_{\rm AC}(T)$, cf.\ Figs.\ \ref{M-ac}A and B, respectively.

Concerning the $^{174}$\yrs sample, we believe that (i) it contains a
low, but finite, concentration of residual Yb ions with nuclear spins which
give rise to a weakened hybrid A-phase order and, thus, weakened
superconductivity, and (ii) \yrs\ free of nuclear spins would not be a
superconductor, at least near $B = 0$. To check this, future
studies of superconductivity of \yrs\ samples with enriched $^{174}$Yb isotope
should be assisted by high-precision mass spectrometry, which is required to
determine the amount of residual nuclear spins.

A three-component Landau theory was applied in \cite{sch16} to explain
the development of two subsequent AF phase transitions at $T_{\rm N}$ and
$T_{\rm A}$. (Assuming two components, one would obtain only one phase
transition). This theoretical treatment was based on the empirical knowledge
that the 4$f$-electronic spin susceptibility $\chi_{4f}(\bm{Q})$ is highly
anisotropic, exhibiting maxima at two wave vectors, $\bm{Q}_{\rm AF}$ and
$\bm{Q} = 0$, and giving rise to the primary AF order and ferromagnetic
correlations \citep{geg05}, respectively.

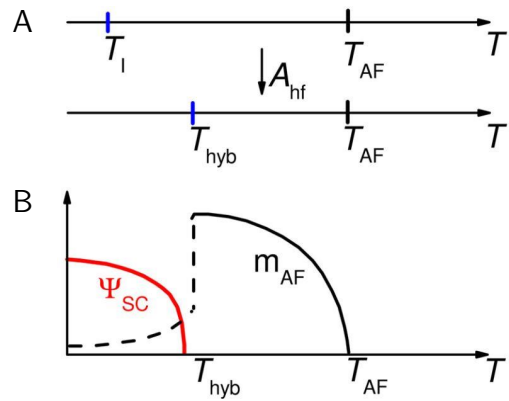
\begin{figure}[t]
\begin{center}
\includegraphics[width=6cm]{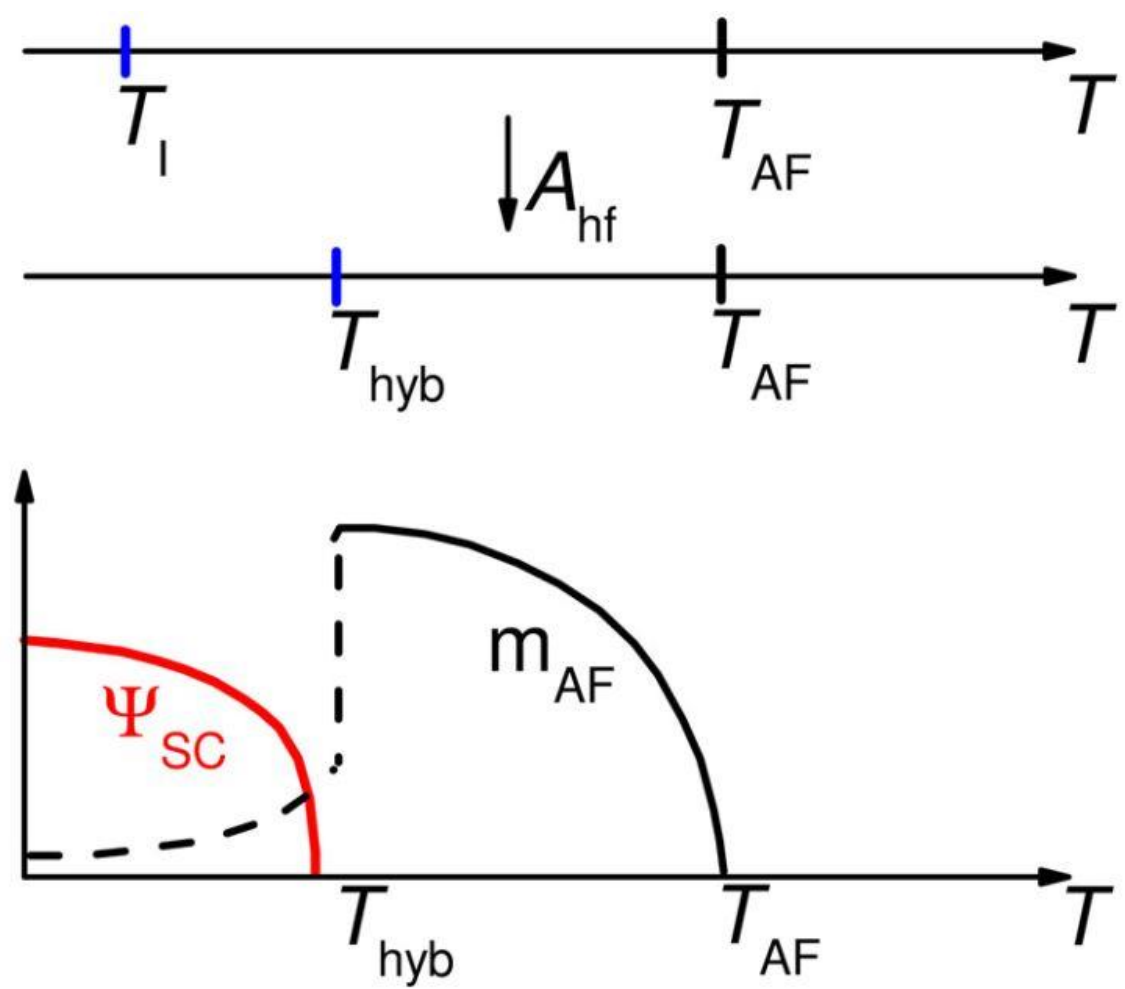}
\end{center}
\unitlength1cm \begin{picture}(-0.4,0.01)
\put(-3.8,3.2){\sffamily\large B}
\put(-3.8,5.6){\sffamily\large A}
\end{picture}
\caption{Three component Landau theory: Phase transitions at $T_{\rm AF}$
and $T_{\rm hyb}$. \textbf{(A)} Sketch of the two phase transitions associated
with electronic and nuclear spin orders. (Top line) Without any hyperfine
coupling ($A_{\rm hf}$), the electronic and nuclear spins are ordered at
$T_{\rm AF}$ and $T_{\rm I}$, respectively. (Bottom line) With hyperfine
coupling, $T_{\rm AF}$ is not affected, but a hybrid nuclear and electronic
spin order is induced at $T_{\rm hyb} \gg T_{\rm I}$. \textbf{(B)} Temperature
evolution of the primary electronic spin order parameter ($m_{\rm AF}$) and
the superconducting order parameter $\Psi_{\rm SC}$. $\Psi_{\rm SC}$ is
developed when $m_{\rm AF}$ is suppressed by the formation of hybrid nuclear
and electronic spin order directly below $T_{\rm hyb}$. Reproduced from
\cite{sch16}.}  \label{GL}
\end{figure}

It is, therefore, natural to assume that a peak in $\chi_{4f}(\bm{Q})$ exists
at yet another finite wave vector ${\bm{Q}_1}$, different from
$\bm{Q}_{\rm AF}$. Along ${\bm{Q}_1}$, the RKKY interaction generates an order
parameter $\Phi_{\rm J}$ among the 4$f$-electronic spins and simultaneously an
order parameter $\Phi_{\rm I}$ among the Yb-derived nuclear spins. According
to the size of the effective $g$-factor discussed above, $\Phi_{\rm I}$ must
be much larger than $\Phi_{\rm J}$. As $\Phi_{\rm J}$ and $\Phi_{\rm I}$ exist
at the same wave vector ${\bm{Q}_1}$, they are coupled bilinearly via $-
\lambda \, \Phi_{\rm J} \Phi_{\rm I}$, where the coupling parameter
$\lambda \approx 25$~mK is related to the hyperfine coupling constant $A_{\rm
hf}$ by $\lambda \sim A_{\rm hf} \approx 100$~T/$\mu_{\rm B}$. As a result,
one finds the transition temperature of the
nuclear-dominated hybrid A-phase $T_{\rm hyb} \approx \lambda^2
\chi_{4f} ({\bm{Q}_1}) / g_{4f}^2 \approx 1$~mK. Here, the value of the
bulk susceptibility was taken for that of the unknown $\chi_{4f} ({\bm{Q}_1})$.
In view of this uncertainty, the agreement between the theoretical result for
$T_{\rm hyb}$ and the experimental value $T_{\rm A} = 2$~mK is surprisingly
good. Within the afore-described Landau treatment it may be assumed that
the nuclear order parameter $\Phi_{\rm I}$ competes with the primary
4$f$-electronic order parameter $\Phi_{\rm AF}$ that was found to be
detrimental to superconductivity. Consequently, superconductivity sets in once
the primary order $\Phi_{\rm AF}$ is suppressed, cf. Figure \ref{GL}.

As this Landau theory is a mean-field theory, it does not treat short-range
order. However, unique nuclear short-range order, as visualized by a
significant temperature dependence of the nuclear spin entropy (Fig.\
\ref{SpecHeat}B), apparently exists up to at least 16~mK and generates the
nucleation of superconductivity at $T \simeq 10$~mK (Fig.\ \ref{M-ac}A). In
reality, as was inferred from Fig.\ \ref{MdcTB}A, the staggered magnetization
of the primary AF phase, $m_{\rm AF}$, starts to decrease already at 20 mK,
rather than at $T = T_{\rm hyb}$. As suggested in Fig.\ \ref{GL},
$m_{\rm AF}(T)$ continues to decrease below $T_{\rm hyb}$ and may well vanish
as $T \rightarrow 0$. This means, the QCP is located at (or very close to)
zero external magnetic field. Interestingly, the transition from the primary
AF order to superconductivity is of first order \citep{sch16}, similar to what
was found for single-crystalline $A/S$-type CeCu$_2$Si$_2$ \citep{sto06}.

In the following we summarize, and offer a---possibly
oversimplified---explanation of, the multitude of puzzling results achieved by
the previous magnetic and calorimetric as well as the recent resistive
measurements:
\renewcommand{\labelenumi}{\roman{enumi}.}
\begin{enumerate}
\item Superconductivity in \yrs\ apparently exists in a wide field range
almost up to $B_{\rm A} = 45$~mT, with the maximum $T_{\rm c}$ occurring
at $B = 0$ \citep{sch16}.
\item The $T$-dependence of fc-$M_{\rm DC}(T)$ displayed in Fig.\
\ref{MdcTB}A demonstrates that the staggered moment, $m_{\rm AF}$, of the
primary order starts to decrease already at about 20~mK, which is close to the
temperature associated with the strength of the local Yb hyperfine interaction
(25 mK). In addition, a significant $T$-dependence of the Yb-derived nuclear
spin entropy, $S_{\rm I,Yb}(T)$, is observed up to $T_{\rm B} \approx$ 16~mK,
where a distinct anomaly is visible in Fig.\ \ref{SpecHeat}. The latter is
associated with a rather sharp removal of a tiny fraction of about 1.5\%
of $S_{\rm tot}$, the full value of $S_{\rm I,Yb}(T)$ at high temperatures.
This can be ascribed to a corresponding small fraction of the Yb-derived
nuclear spins which interact with the mean hyperfine field induced by the
4$f$-electron spins \citep{sch16} to create random and insulated A-phase
regions which must be growing upon lowering the temperature.
\item The simultaneous onset of both a weak, but distinct shielding
signal (Fig.\ \ref{M-ac}) and a pronounced decline in the absolute slope of
fc $M_{\rm DC}(T)$ (Fig.\ \ref{MdcTB}A) at $T \approx 10$~mK are
interpreted as manifesting the nucleation of superconductivity.
\item The network of superconducting islands turns out to be extremely
fragile as the shielding signal is suppressed already by a very small external
magnetic field (Fig.\ \ref{M-ac}A).
\item Upon further cooling, the resistivity reaches zero at about 6.5~mK.
This can be reconciled with the above results in terms of the onset
of superconducting percolation  as mentioned before. The percolation
initiates a considerably increasing strength of A-phase
short-range order as illustrated by a pronounced increase of fc-$M_{\rm DC}(T)$
below 5 mK (Fig.\ \ref{MdcTB}A) and corroborated by a significant removal of
the Yb-derived nuclear spin entropy from its maximum value $S_{\rm tot}$ (at
$T > T_{\rm B}$) by 26\% on cooling the sample down to $T_{\rm A}$ (Fig.\
\ref{SpecHeat}B). Remarkably, the shielding signal due to the network of
superconducting islands in the sample surface does not exceed 20\% when
cooling from $T \approx 10$~mK to $T \ge T_{\rm A}$, the temperature range
previously called `B-phase', cf.\ Fig.\ \ref{lowTphase}.
\item At $T_{\rm A}$ ($= 2.3$~mK as $B \rightarrow 0$, see \cite{sch16}), a
second-order phase transition into long-range A-phase order takes place (Fig.\
\ref{SpecHeat}A). The ordering of the Yb nuclear spins reaches a level of
more than 60\% already at 1.5~mK, see Fig.\ \ref{SpecHeat}B.
Nuclear-dominated hybrid order at elevated magnetic fields has yet to be
confirmed by future measurements of the specific heat. This should clarify
whether the A-phase exists up to $B_{\rm A} \simeq 45$~mT, as indicated by the
dotted red line in the phase diagram of Fig.\ \ref{lowTphase}, or even up to
the quantum critical field $B_{\rm N} = 60$~mT, as suggested by the new
resistivity results of \cite{ngu21} (Fig.\ \ref{rhoN}). In the latter
case, we conjecture the existence of not just two QCPs at $B = B_{\rm N}$
(vanishing of the primary AF order due to the applied magnetic field) and at
$B = 0$ (due to the competing A-phase). Rather, a phase boundary $T_{\rm A}(B)$
spanning the whole field range $B < B_{\rm N}$ would exist and most likely
establish a \emph{quantum critical line} between these two QCPs.
\item From the magnetization measurements \citep{sch16}, bulk
superconductivity emerges at $T_{\rm c} = 2$~mK (Figs.\ \ref{MdcTB}A and
\ref{M-ac}A), slightly below $T_{\rm A}$. Future fc-$M_{\rm DC}(T)$
measurements on annealed powder samples below the lower critical field
$B_{c1}$ are needed to confirm that bulk superconductivity indeed exists well
below $T_{c,\rho}$ at magnetic fields up to $B \simeq B_{\rm A}$.
\end{enumerate}

We should welcome future work by other groups to cross-check the above
reasoning.

The new resistivity study by \cite{ngu21} suggests the existence of two
separate superconducting regimes, at low and elevated fields, respectively.
In view of the preceding discussion, we cannot concur with the proposal by
Nguyen et al.\ that, at least in the low-field regime, Cooper pairing in \yrs\
is mediated by the critical fluctuations of the \emph{field-induced} QCP
at $B_{\rm N}$: For, here the primary AF order is apparently suppressed by the
competing nuclear order, allowing a quantum critical line to be established.
Therefore, it is natural to consider the associated quantum critical
fluctuations at \emph{very low fields} being the driving force for the
formation of Cooper pairs.

The origin of the superconductivity at elevated fields remains to be
resolved. If here, the existence of the nuclear-dominated hybrid A-phase can
be proven by future specific-heat experiments, the Cooper-pairing mechanism
emphasized above in the low-field regime will most likely be operating as well.
An alternative scenario proposed by \cite{ngu21} neglects the competition
between the nuclear and primary 4$f$-electronic orders and instead implies, as
mentioned before, that the quantum critical fluctuations associated with the
\emph{field-induced} QCP at $B_{\rm N} = 60$~mT bring about the formation
of superconductivity. As these critical fluctuations would be strongly impeded
by the pair-breaking applied magnetic field, the competition between the
field-induced quantum criticality and the destructive action of the magnetic
field on the Cooper pairs causes a maximum $T_{\rm c}$ to occur well
below $B_{\rm N}$. \cite{ngu21} propose that
superconductivity at elevated fields may be of the spin-triplet variety. While
this would not have been a surprise had the quantum criticality been of nearly
ferromagnetic (and conventional) type \citep{li19}, how could spin-triplet
pairing develop for Kondo-destroying type AF quantum criticality as is the
case in \yrs? Insights into this question have come from theoretical studies
near a Kondo-destroying QCP in antiferromagnetically coupled Kondo models
\citep{hu21}: They showed that significant Ising anisotropy, as it effectively
arises under an external magnetic field in the easy-plane antiferromagnetically
correlated \yrs, makes the spin-triplet pairing competitive.

\section{Perspective}
The discovery of superconductivity in \yrs\ at ultra-low temperatures opens up
new dimensions. It expands the horizon of ultra-low temperature physics towards
strongly correlated electronic matter in solids and conversely, it determines
a new area of heavy-fermion physics by reaching down to ultra-low temperatures.

Up to now, heavy-fermion superconductivity has been detected in more than
fifty $f$-electron compounds. Despite the fact that, compared with PuCoGa$_5$, exhibiting the record-high $T_{\rm c}$ of 18.5 K \citep{sar02}, \yrs\ shows a
transition temperature which is lower by almost four decades, this may be
called `high $T_{\rm c}$’---in the sense that $T_{\rm c}$ is limited by an
enormously high ordering temperature of nuclear spins. This nuclear order paves
the way for the superconductivity, not by actively helping in the Cooper-pair
formation, rather by destroying the 4$f$-electronic order below $T_{\rm N} =
70$~mK, which appears to be extremely hostile to superconductivity. The
apparent competition between the nuclear and the 4$f$-electronic orders
results in an AF QCP at, or very close to, $B = 0$. Most likely, the
latter---like its field-induced counterpart at $B = B_{\rm N}$---is of the
`local' type, as inferred from the $B = 0$ results of electrical and thermal
transport as well as specific-heat measurements in the paramagnetic state as
presented in Section \ref{sec1}.

The ultra-low temperature work on \yrs\ strongly supports the notion that
superconductivity robustly develops in the vicinity of such a `partial-Mott'
QCP, as has been theoretically derived in Kondo-lattice models for a
Kondo-destroying QCP \citep{hu21b} and experimentally evidenced from de
Haas-van Alphen studies in high magnetic fields \citep{shi05} and transport
measurements \citep{par06} on pressurized CeRhIn$_5$. Therefore, the results
on these two compounds provide the long-sought \citep{map04} link between
unconventional superconductivity in heavy-fermion metals \citep{pfl09} and
that occurring near a true Mott metal-insulator transition, e.g., in the
cuprates \citep{lee06} and organic charge-transfer salts \citep{kan08}.

\section*{Author Contributions}
All authors contributed substantially to the discussion and revision of
the manuscript.\\

\section*{Funding}
%Work at Rice University was supported in part by the Air Force Office of
%Scientific Research under Grant No.\ FA9550-21-1-0356 and the Robert A. Welch
%Foundation under Grant No.\ C-1411. Q.S. acknowledges the hospitality of the
%Aspen Center for Physics, which is supported by the NSF under Grant No.\
%PHY-1607611.
S.W.\ acknowledges support by the Deutsche Forschungsgemeinschaft through
WI 1324/5-1.

\section*{Conflict of Interest Statement}
The authors declare that the research was conducted in the absence of any
commercial or financial relationships that could be construed as a potential
conflict of interest.

\section*{Acknowledgments}
We are grateful for conversations with Manuel Brando, Ang Cai, Lei Chen, Piers
Coleman, Haoyu Hu, Stefan Lausberg, Silke Paschen, John Saunders, Rong Yu and
in particular with Qimiao Si for his insightful comments.


\begin{thebibliography}{71}
\expandafter\ifx\csname natexlab\endcsname\relax\def\natexlab#1{#1}\fi
\expandafter\ifx\csname bibnamefont\endcsname\relax
  \def\bibnamefont#1{#1}\fi
\expandafter\ifx\csname bibfnamefont\endcsname\relax
  \def\bibfnamefont#1{#1}\fi
\expandafter\ifx\csname citenamefont\endcsname\relax
  \def\citenamefont#1{#1}\fi
\expandafter\ifx\csname url\endcsname\relax
  \def\url#1{\texttt{#1}}\fi
\expandafter\ifx\csname urlprefix\endcsname\relax\def\urlprefix{URL }\fi
\providecommand{\bibinfo}[2]{#2}
\providecommand{\eprint}[2][]{\url{#2}}

\bibitem[{\citenamefont{Schuberth et~al.}(2016)\citenamefont{Schuberth,
  Tippmann, Steinke, Lausberg, Steppke, Brando, Krellner, Geibel, Yu, Si
  et~al.}}]{sch16}
\bibinfo{author}{\bibfnamefont{E.}~\bibnamefont{Schuberth}},
  \bibinfo{author}{\bibfnamefont{M.}~\bibnamefont{Tippmann}},
  \bibinfo{author}{\bibfnamefont{L.}~\bibnamefont{Steinke}},
  \bibinfo{author}{\bibfnamefont{S.}~\bibnamefont{Lausberg}},
  \bibinfo{author}{\bibfnamefont{A.}~\bibnamefont{Steppke}},
  \bibinfo{author}{\bibfnamefont{M.}~\bibnamefont{Brando}},
  \bibinfo{author}{\bibfnamefont{C.}~\bibnamefont{Krellner}},
  \bibinfo{author}{\bibfnamefont{C.}~\bibnamefont{Geibel}},
  \bibinfo{author}{\bibfnamefont{R.}~\bibnamefont{Yu}},
  \bibinfo{author}{\bibfnamefont{Q.}~\bibnamefont{Si}}, \bibnamefont{et~al.},
  \bibinfo{journal}{Science} \textbf{\bibinfo{volume}{351}},
  \bibinfo{pages}{485} (\bibinfo{year}{2016}).

\bibitem[{\citenamefont{Nguyen et~al.}(2021)\citenamefont{Nguyen, Sidorenko,
  Taupin, Knebel, Lapertot, Schuberth, and Paschen}}]{ngu21}
\bibinfo{author}{\bibfnamefont{D.~H.} \bibnamefont{Nguyen}},
  \bibinfo{author}{\bibfnamefont{A.}~\bibnamefont{Sidorenko}},
  \bibinfo{author}{\bibfnamefont{M.}~\bibnamefont{Taupin}},
  \bibinfo{author}{\bibfnamefont{G.}~\bibnamefont{Knebel}},
  \bibinfo{author}{\bibfnamefont{G.}~\bibnamefont{Lapertot}},
  \bibinfo{author}{\bibfnamefont{E.}~\bibnamefont{Schuberth}},
  \bibnamefont{and} \bibinfo{author}{\bibfnamefont{S.}~\bibnamefont{Paschen}},
  \bibinfo{journal}{Nat.\ Commun.} \textbf{\bibinfo{volume}{12}},
  \bibinfo{pages}{4341} (\bibinfo{year}{2021}).

\bibitem[{\citenamefont{Wirth and Steglich}(2016)}]{wir16}
\bibinfo{author}{\bibfnamefont{S.}~\bibnamefont{Wirth}} \bibnamefont{and}
  \bibinfo{author}{\bibfnamefont{F.}~\bibnamefont{Steglich}},
  \bibinfo{journal}{Nat.\ Rev.\ Mat.} \textbf{\bibinfo{volume}{1}},
  \bibinfo{pages}{16051} (\bibinfo{year}{2016}).

\bibitem[{\citenamefont{Chen et~al.}(2017)\citenamefont{Chen, Xu, Niu, Jiang,
  Peng, Xu, Wen, Ding, Huang, Shu et~al.}}]{che17}
\bibinfo{author}{\bibfnamefont{Q.~Y.} \bibnamefont{Chen}},
  \bibinfo{author}{\bibfnamefont{D.~F.} \bibnamefont{Xu}},
  \bibinfo{author}{\bibfnamefont{X.~H.} \bibnamefont{Niu}},
  \bibinfo{author}{\bibfnamefont{J.}~\bibnamefont{Jiang}},
  \bibinfo{author}{\bibfnamefont{R.}~\bibnamefont{Peng}},
  \bibinfo{author}{\bibfnamefont{H.~C.} \bibnamefont{Xu}},
  \bibinfo{author}{\bibfnamefont{C.~H.~P.} \bibnamefont{Wen}},
  \bibinfo{author}{\bibfnamefont{Z.~F.} \bibnamefont{Ding}},
  \bibinfo{author}{\bibfnamefont{K.}~\bibnamefont{Huang}},
  \bibinfo{author}{\bibfnamefont{L.}~\bibnamefont{Shu}}, \bibnamefont{et~al.},
  \bibinfo{journal}{Phys.\ Rev.\ B} \textbf{\bibinfo{volume}{96}},
  \bibinfo{pages}{045107} (\bibinfo{year}{2017}).

\bibitem[{\citenamefont{Ernst et~al.}(2011)\citenamefont{Ernst, Kirchner,
  Krellner, Geibel, Zwicknagl, Steglich, and Wirth}}]{ern11}
\bibinfo{author}{\bibfnamefont{S.}~\bibnamefont{Ernst}},
  \bibinfo{author}{\bibfnamefont{S.}~\bibnamefont{Kirchner}},
  \bibinfo{author}{\bibfnamefont{C.}~\bibnamefont{Krellner}},
  \bibinfo{author}{\bibfnamefont{C.}~\bibnamefont{Geibel}},
  \bibinfo{author}{\bibfnamefont{G.}~\bibnamefont{Zwicknagl}},
  \bibinfo{author}{\bibfnamefont{F.}~\bibnamefont{Steglich}}, \bibnamefont{and}
  \bibinfo{author}{\bibfnamefont{S.}~\bibnamefont{Wirth}},
  \bibinfo{journal}{Nature} \textbf{\bibinfo{volume}{474}},
  \bibinfo{pages}{362} (\bibinfo{year}{2011}).

\bibitem[{\citenamefont{Seiro et~al.}(2018)\citenamefont{Seiro, Jiao, Kirchner,
  Hartmann, Friedemann, Krellner, Geibel, Si, Steglich, and Wirth}}]{sei18}
\bibinfo{author}{\bibfnamefont{S.}~\bibnamefont{Seiro}},
  \bibinfo{author}{\bibfnamefont{L.}~\bibnamefont{Jiao}},
  \bibinfo{author}{\bibfnamefont{S.}~\bibnamefont{Kirchner}},
  \bibinfo{author}{\bibfnamefont{S.}~\bibnamefont{Hartmann}},
  \bibinfo{author}{\bibfnamefont{S.}~\bibnamefont{Friedemann}},
  \bibinfo{author}{\bibfnamefont{C.}~\bibnamefont{Krellner}},
  \bibinfo{author}{\bibfnamefont{C.}~\bibnamefont{Geibel}},
  \bibinfo{author}{\bibfnamefont{Q.}~\bibnamefont{Si}},
  \bibinfo{author}{\bibfnamefont{F.}~\bibnamefont{Steglich}}, \bibnamefont{and}
  \bibinfo{author}{\bibfnamefont{S.}~\bibnamefont{Wirth}},
  \bibinfo{journal}{Nat.\ Commun.} \textbf{\bibinfo{volume}{9}},
  \bibinfo{pages}{3324} (\bibinfo{year}{2018}).

\bibitem[{\citenamefont{Stewart}(2001)}]{stew01}
\bibinfo{author}{\bibfnamefont{G.~R.} \bibnamefont{Stewart}},
  \bibinfo{journal}{Rev.\ Mod.\ Phys.} \textbf{\bibinfo{volume}{73}},
  \bibinfo{pages}{797} (\bibinfo{year}{2001}), \bibinfo{note}{{\it ibid.} {\bf
  78}, 743-753 (2006)}.

\bibitem[{\citenamefont{{von L{\"o}hneysen} et~al.}(2007)\citenamefont{{von
  L{\"o}hneysen}, Rosch, Vojta, and W{\"o}lfle}}]{hvl07}
\bibinfo{author}{\bibfnamefont{H.}~\bibnamefont{{von L{\"o}hneysen}}},
  \bibinfo{author}{\bibfnamefont{A.}~\bibnamefont{Rosch}},
  \bibinfo{author}{\bibfnamefont{M.}~\bibnamefont{Vojta}}, \bibnamefont{and}
  \bibinfo{author}{\bibfnamefont{P.}~\bibnamefont{W{\"o}lfle}},
  \bibinfo{journal}{Rev.\ Mod.\ Phys.} \textbf{\bibinfo{volume}{79}},
  \bibinfo{pages}{1015} (\bibinfo{year}{2007}).

\bibitem[{\citenamefont{Gegenwart et~al.}(2008)\citenamefont{Gegenwart,
  Steglich, and Si}}]{geg08}
\bibinfo{author}{\bibfnamefont{P.}~\bibnamefont{Gegenwart}},
  \bibinfo{author}{\bibfnamefont{F.}~\bibnamefont{Steglich}}, \bibnamefont{and}
  \bibinfo{author}{\bibfnamefont{Q.}~\bibnamefont{Si}},
  \bibinfo{journal}{Nature Phys.} \textbf{\bibinfo{volume}{4}},
  \bibinfo{pages}{186} (\bibinfo{year}{2008}).

\bibitem[{\citenamefont{Sachdev}(2011)}]{sac11}
\bibinfo{author}{\bibfnamefont{S.}~\bibnamefont{Sachdev}},
  \emph{\bibinfo{title}{Quantum phase transitions}}
  (\bibinfo{publisher}{Cambridge University Press}, \bibinfo{year}{2011}).

\bibitem[{\citenamefont{K{\"o}hler et~al.}(2008)\citenamefont{K{\"o}hler,
  Oeschler, Steglich, Maquilon, and Fisk}}]{koe08}
\bibinfo{author}{\bibfnamefont{U.}~\bibnamefont{K{\"o}hler}},
  \bibinfo{author}{\bibfnamefont{N.}~\bibnamefont{Oeschler}},
  \bibinfo{author}{\bibfnamefont{F.}~\bibnamefont{Steglich}},
  \bibinfo{author}{\bibfnamefont{S.}~\bibnamefont{Maquilon}}, \bibnamefont{and}
  \bibinfo{author}{\bibfnamefont{Z.}~\bibnamefont{Fisk}},
  \bibinfo{journal}{Phys.\ Rev.\ B} \textbf{\bibinfo{volume}{77}},
  \bibinfo{pages}{104412} (\bibinfo{year}{2008}).

\bibitem[{\citenamefont{Trovarelli et~al.}(2000)\citenamefont{Trovarelli,
  Geibel, Mederle, Langhammer, Grosche, Gegenwart, Lang, Sparn, and
  Steglich}}]{tro00}
\bibinfo{author}{\bibfnamefont{O.}~\bibnamefont{Trovarelli}},
  \bibinfo{author}{\bibfnamefont{C.}~\bibnamefont{Geibel}},
  \bibinfo{author}{\bibfnamefont{S.}~\bibnamefont{Mederle}},
  \bibinfo{author}{\bibfnamefont{C.}~\bibnamefont{Langhammer}},
  \bibinfo{author}{\bibfnamefont{F.~M.} \bibnamefont{Grosche}},
  \bibinfo{author}{\bibfnamefont{P.}~\bibnamefont{Gegenwart}},
  \bibinfo{author}{\bibfnamefont{M.}~\bibnamefont{Lang}},
  \bibinfo{author}{\bibfnamefont{G.}~\bibnamefont{Sparn}}, \bibnamefont{and}
  \bibinfo{author}{\bibfnamefont{F.}~\bibnamefont{Steglich}},
  \bibinfo{journal}{Phys.\ Rev.\ Lett.} \textbf{\bibinfo{volume}{85}},
  \bibinfo{pages}{626} (\bibinfo{year}{2000}).

\bibitem[{\citenamefont{Gegenwart et~al.}(2002)\citenamefont{Gegenwart,
  Custers, Geibel, Neumaier, Tayama, Tenya, Trovarelli, and Steglich}}]{geg02}
\bibinfo{author}{\bibfnamefont{P.}~\bibnamefont{Gegenwart}},
  \bibinfo{author}{\bibfnamefont{J.}~\bibnamefont{Custers}},
  \bibinfo{author}{\bibfnamefont{C.}~\bibnamefont{Geibel}},
  \bibinfo{author}{\bibfnamefont{K.}~\bibnamefont{Neumaier}},
  \bibinfo{author}{\bibfnamefont{T.}~\bibnamefont{Tayama}},
  \bibinfo{author}{\bibfnamefont{K.}~\bibnamefont{Tenya}},
  \bibinfo{author}{\bibfnamefont{O.}~\bibnamefont{Trovarelli}},
  \bibnamefont{and} \bibinfo{author}{\bibfnamefont{F.}~\bibnamefont{Steglich}},
  \bibinfo{journal}{Phys.\ Rev.\ Lett.} \textbf{\bibinfo{volume}{89}},
  \bibinfo{pages}{056402} (\bibinfo{year}{2002}).

\bibitem[{\citenamefont{Ishida et~al.}(2003)\citenamefont{Ishida, MacLaughlin,
  Young, Okamoto, Kawasaki, Kitaoka, Nieuwenhuys, Heffner, Bernal, Higemoto
  et~al.}}]{ish03}
\bibinfo{author}{\bibfnamefont{K.}~\bibnamefont{Ishida}},
  \bibinfo{author}{\bibfnamefont{D.~E.} \bibnamefont{MacLaughlin}},
  \bibinfo{author}{\bibfnamefont{B.-L.} \bibnamefont{Young}},
  \bibinfo{author}{\bibfnamefont{K.}~\bibnamefont{Okamoto}},
  \bibinfo{author}{\bibfnamefont{Y.}~\bibnamefont{Kawasaki}},
  \bibinfo{author}{\bibfnamefont{Y.}~\bibnamefont{Kitaoka}},
  \bibinfo{author}{\bibfnamefont{G.~J.} \bibnamefont{Nieuwenhuys}},
  \bibinfo{author}{\bibfnamefont{R.~H.} \bibnamefont{Heffner}},
  \bibinfo{author}{\bibfnamefont{O.~O.} \bibnamefont{Bernal}},
  \bibinfo{author}{\bibfnamefont{W.}~\bibnamefont{Higemoto}},
  \bibnamefont{et~al.}, \bibinfo{journal}{Phys.\ Rev.\ B}
  \textbf{\bibinfo{volume}{68}}, \bibinfo{pages}{184401}
  (\bibinfo{year}{2003}).

\bibitem[{\citenamefont{Gegenwart et~al.}(2007)\citenamefont{Gegenwart,
  Westerkamp, Krellner, Tokiwa, Paschen, Geibel, Steglich, Abrahams, and
  Si}}]{geg07}
\bibinfo{author}{\bibfnamefont{P.}~\bibnamefont{Gegenwart}},
  \bibinfo{author}{\bibfnamefont{T.}~\bibnamefont{Westerkamp}},
  \bibinfo{author}{\bibfnamefont{C.}~\bibnamefont{Krellner}},
  \bibinfo{author}{\bibfnamefont{Y.}~\bibnamefont{Tokiwa}},
  \bibinfo{author}{\bibfnamefont{S.}~\bibnamefont{Paschen}},
  \bibinfo{author}{\bibfnamefont{C.}~\bibnamefont{Geibel}},
  \bibinfo{author}{\bibfnamefont{F.}~\bibnamefont{Steglich}},
  \bibinfo{author}{\bibfnamefont{E.}~\bibnamefont{Abrahams}}, \bibnamefont{and}
  \bibinfo{author}{\bibfnamefont{Q.}~\bibnamefont{Si}},
  \bibinfo{journal}{Science} \textbf{\bibinfo{volume}{315}},
  \bibinfo{pages}{969} (\bibinfo{year}{2007}).

\bibitem[{\citenamefont{Paschen et~al.}(2004)\citenamefont{Paschen,
  L{\"u}hmann, Wirth, Gegenwart, Trovarelli, Geibel, Steglich, Coleman, and
  Si}}]{pas04}
\bibinfo{author}{\bibfnamefont{S.}~\bibnamefont{Paschen}},
  \bibinfo{author}{\bibfnamefont{T.}~\bibnamefont{L{\"u}hmann}},
  \bibinfo{author}{\bibfnamefont{S.}~\bibnamefont{Wirth}},
  \bibinfo{author}{\bibfnamefont{P.}~\bibnamefont{Gegenwart}},
  \bibinfo{author}{\bibfnamefont{O.}~\bibnamefont{Trovarelli}},
  \bibinfo{author}{\bibfnamefont{C.}~\bibnamefont{Geibel}},
  \bibinfo{author}{\bibfnamefont{F.}~\bibnamefont{Steglich}},
  \bibinfo{author}{\bibfnamefont{P.}~\bibnamefont{Coleman}}, \bibnamefont{and}
  \bibinfo{author}{\bibfnamefont{Q.}~\bibnamefont{Si}},
  \bibinfo{journal}{Nature} \textbf{\bibinfo{volume}{432}},
  \bibinfo{pages}{881} (\bibinfo{year}{2004}).

\bibitem[{\citenamefont{Friedemann et~al.}(2010)\citenamefont{Friedemann,
  Oeschler, Wirth, Krellner, Geibel, Steglich, Paschen, Kirchner, and
  Si}}]{Fri10}
\bibinfo{author}{\bibfnamefont{S.}~\bibnamefont{Friedemann}},
  \bibinfo{author}{\bibfnamefont{N.}~\bibnamefont{Oeschler}},
  \bibinfo{author}{\bibfnamefont{S.}~\bibnamefont{Wirth}},
  \bibinfo{author}{\bibfnamefont{C.}~\bibnamefont{Krellner}},
  \bibinfo{author}{\bibfnamefont{C.}~\bibnamefont{Geibel}},
  \bibinfo{author}{\bibfnamefont{F.}~\bibnamefont{Steglich}},
  \bibinfo{author}{\bibfnamefont{S.}~\bibnamefont{Paschen}},
  \bibinfo{author}{\bibfnamefont{S.}~\bibnamefont{Kirchner}}, \bibnamefont{and}
  \bibinfo{author}{\bibfnamefont{Q.}~\bibnamefont{Si}},
  \bibinfo{journal}{Proc.\ Natl.\ Acad.\ Sci.\ USA}
  \textbf{\bibinfo{volume}{107}}, \bibinfo{pages}{14547}
  (\bibinfo{year}{2010}).

\bibitem[{\citenamefont{Si et~al.}(2001)\citenamefont{Si, Rabello, Ingersent,
  and Smith}}]{qsi01}
\bibinfo{author}{\bibfnamefont{Q.}~\bibnamefont{Si}},
  \bibinfo{author}{\bibfnamefont{S.}~\bibnamefont{Rabello}},
  \bibinfo{author}{\bibfnamefont{K.}~\bibnamefont{Ingersent}},
  \bibnamefont{and} \bibinfo{author}{\bibfnamefont{J.~L.} \bibnamefont{Smith}},
  \bibinfo{journal}{Nature} \textbf{\bibinfo{volume}{413}},
  \bibinfo{pages}{804} (\bibinfo{year}{2001}).

\bibitem[{\citenamefont{Coleman et~al.}(2001)\citenamefont{Coleman, P{\'e}pin,
  Si, and Ramazashvili}}]{col01}
\bibinfo{author}{\bibfnamefont{P.}~\bibnamefont{Coleman}},
  \bibinfo{author}{\bibfnamefont{C.}~\bibnamefont{P{\'e}pin}},
  \bibinfo{author}{\bibfnamefont{Q.}~\bibnamefont{Si}}, \bibnamefont{and}
  \bibinfo{author}{\bibfnamefont{R.}~\bibnamefont{Ramazashvili}},
  \bibinfo{journal}{J.\ Phys.: Condens.\ Matter} \textbf{\bibinfo{volume}{13}},
  \bibinfo{pages}{R723} (\bibinfo{year}{2001}).

\bibitem[{\citenamefont{Aronson et~al.}(1995)\citenamefont{Aronson, Osborn,
  Robinson, Lynn, Chau, Seaman, and Maple}}]{aro95}
\bibinfo{author}{\bibfnamefont{M.~C.} \bibnamefont{Aronson}},
  \bibinfo{author}{\bibfnamefont{R.}~\bibnamefont{Osborn}},
  \bibinfo{author}{\bibfnamefont{R.~A.} \bibnamefont{Robinson}},
  \bibinfo{author}{\bibfnamefont{J.~W.} \bibnamefont{Lynn}},
  \bibinfo{author}{\bibfnamefont{R.}~\bibnamefont{Chau}},
  \bibinfo{author}{\bibfnamefont{C.~L.} \bibnamefont{Seaman}},
  \bibnamefont{and} \bibinfo{author}{\bibfnamefont{M.~B.} \bibnamefont{Maple}},
  \bibinfo{journal}{Phys.\ Rev.\ Lett.} \textbf{\bibinfo{volume}{75}},
  \bibinfo{pages}{725} (\bibinfo{year}{1995}).

\bibitem[{\citenamefont{Schr{\"o}der et~al.}(2000)\citenamefont{Schr{\"o}der,
  Aeppli, Coldea, Adams, Stockert, {von L{\"o}hneysen}, Bucher, Ramazashvili,
  and Coleman}}]{loe00}
\bibinfo{author}{\bibfnamefont{A.}~\bibnamefont{Schr{\"o}der}},
  \bibinfo{author}{\bibfnamefont{G.}~\bibnamefont{Aeppli}},
  \bibinfo{author}{\bibfnamefont{R.}~\bibnamefont{Coldea}},
  \bibinfo{author}{\bibfnamefont{M.}~\bibnamefont{Adams}},
  \bibinfo{author}{\bibfnamefont{O.}~\bibnamefont{Stockert}},
  \bibinfo{author}{\bibfnamefont{H.}~\bibnamefont{{von L{\"o}hneysen}}},
  \bibinfo{author}{\bibfnamefont{E.}~\bibnamefont{Bucher}},
  \bibinfo{author}{\bibfnamefont{R.}~\bibnamefont{Ramazashvili}},
  \bibnamefont{and} \bibinfo{author}{\bibfnamefont{P.}~\bibnamefont{Coleman}},
  \bibinfo{journal}{Nature} \textbf{\bibinfo{volume}{407}},
  \bibinfo{pages}{351} (\bibinfo{year}{2000}).

\bibitem[{\citenamefont{{von L{\"o}hneysen} et~al.}(1994)\citenamefont{{von
  L{\"o}hneysen}, Pietrus, Portisch, Schlager, Schr{\"o}der, Sieck, and
  Trappmann}}]{hvl94}
\bibinfo{author}{\bibfnamefont{H.}~\bibnamefont{{von L{\"o}hneysen}}},
  \bibinfo{author}{\bibfnamefont{T.}~\bibnamefont{Pietrus}},
  \bibinfo{author}{\bibfnamefont{G.}~\bibnamefont{Portisch}},
  \bibinfo{author}{\bibfnamefont{H.~G.} \bibnamefont{Schlager}},
  \bibinfo{author}{\bibfnamefont{A.}~\bibnamefont{Schr{\"o}der}},
  \bibinfo{author}{\bibfnamefont{M.}~\bibnamefont{Sieck}}, \bibnamefont{and}
  \bibinfo{author}{\bibfnamefont{T.}~\bibnamefont{Trappmann}},
  \bibinfo{journal}{Phys.\ Rev.\ Lett.} \textbf{\bibinfo{volume}{72}},
  \bibinfo{pages}{3262} (\bibinfo{year}{1994}).

\bibitem[{\citenamefont{Shishido et~al.}(2005)\citenamefont{Shishido, Settai,
  Harima, and \={O}nuki}}]{shi05}
\bibinfo{author}{\bibfnamefont{H.}~\bibnamefont{Shishido}},
  \bibinfo{author}{\bibfnamefont{R.}~\bibnamefont{Settai}},
  \bibinfo{author}{\bibfnamefont{H.}~\bibnamefont{Harima}}, \bibnamefont{and}
  \bibinfo{author}{\bibfnamefont{Y.}~\bibnamefont{\={O}nuki}},
  \bibinfo{journal}{J. Phys. Soc. Jpn.} \textbf{\bibinfo{volume}{74}},
  \bibinfo{pages}{1103} (\bibinfo{year}{2005}).

\bibitem[{\citenamefont{Park et~al.}(2006)\citenamefont{Park, Ronning, Yuan,
  Salamon, Movshovich, Sarrao, and Thompson}}]{par06}
\bibinfo{author}{\bibfnamefont{T.}~\bibnamefont{Park}},
  \bibinfo{author}{\bibfnamefont{F.}~\bibnamefont{Ronning}},
  \bibinfo{author}{\bibfnamefont{H.~Q.} \bibnamefont{Yuan}},
  \bibinfo{author}{\bibfnamefont{M.~B.} \bibnamefont{Salamon}},
  \bibinfo{author}{\bibfnamefont{R.}~\bibnamefont{Movshovich}},
  \bibinfo{author}{\bibfnamefont{J.~L.} \bibnamefont{Sarrao}},
  \bibnamefont{and} \bibinfo{author}{\bibfnamefont{J.~D.}
  \bibnamefont{Thompson}}, \bibinfo{journal}{Nature}
  \textbf{\bibinfo{volume}{440}}, \bibinfo{pages}{65} (\bibinfo{year}{2006}).

\bibitem[{\citenamefont{Custers et~al.}(2003)\citenamefont{Custers, Gegenwart,
  Wilhelm, Neumaier, Tokiwa, Trovarelli, Geibel, Steglich, P{\'e}pin, and
  Coleman}}]{cus03}
\bibinfo{author}{\bibfnamefont{J.}~\bibnamefont{Custers}},
  \bibinfo{author}{\bibfnamefont{P.}~\bibnamefont{Gegenwart}},
  \bibinfo{author}{\bibfnamefont{H.}~\bibnamefont{Wilhelm}},
  \bibinfo{author}{\bibfnamefont{K.}~\bibnamefont{Neumaier}},
  \bibinfo{author}{\bibfnamefont{Y.}~\bibnamefont{Tokiwa}},
  \bibinfo{author}{\bibfnamefont{O.}~\bibnamefont{Trovarelli}},
  \bibinfo{author}{\bibfnamefont{C.}~\bibnamefont{Geibel}},
  \bibinfo{author}{\bibfnamefont{F.}~\bibnamefont{Steglich}},
  \bibinfo{author}{\bibfnamefont{C.}~\bibnamefont{P{\'e}pin}},
  \bibnamefont{and} \bibinfo{author}{\bibfnamefont{P.}~\bibnamefont{Coleman}},
  \bibinfo{journal}{Nature} \textbf{\bibinfo{volume}{424}},
  \bibinfo{pages}{524} (\bibinfo{year}{2003}).

\bibitem[{\citenamefont{Westerkamp}(2009)}]{wes09}
\bibinfo{author}{\bibfnamefont{T.}~\bibnamefont{Westerkamp}}, Ph.D. thesis,
  \bibinfo{school}{Technical University Dresden, Germany}
  (\bibinfo{year}{2009}).

\bibitem[{\citenamefont{W{\"o}lfle and Abrahams}(2011)}]{woe11}
\bibinfo{author}{\bibfnamefont{P.}~\bibnamefont{W{\"o}lfle}} \bibnamefont{and}
  \bibinfo{author}{\bibfnamefont{E.}~\bibnamefont{Abrahams}},
  \bibinfo{journal}{Phys.\ Rev.\ B} \textbf{\bibinfo{volume}{84}},
  \bibinfo{pages}{041101(R)} (\bibinfo{year}{2011}).

\bibitem[{\citenamefont{Taupin and Paschen}(2022)}]{tau22}
\bibinfo{author}{\bibfnamefont{M.}~\bibnamefont{Taupin}} \bibnamefont{and}
  \bibinfo{author}{\bibfnamefont{S.}~\bibnamefont{Paschen}}
  (\bibinfo{year}{2022}), \bibinfo{note}{\textit{{a}rXiv}:2201.02820}.

\bibitem[{\citenamefont{Kirchner et~al.}(2020)\citenamefont{Kirchner, Paschen,
  Chen, Wirth, Feng, Thompson, and Si}}]{kir20}
\bibinfo{author}{\bibfnamefont{S.}~\bibnamefont{Kirchner}},
  \bibinfo{author}{\bibfnamefont{S.}~\bibnamefont{Paschen}},
  \bibinfo{author}{\bibfnamefont{Q.}~\bibnamefont{Chen}},
  \bibinfo{author}{\bibfnamefont{S.}~\bibnamefont{Wirth}},
  \bibinfo{author}{\bibfnamefont{D.}~\bibnamefont{Feng}},
  \bibinfo{author}{\bibfnamefont{J.~D.} \bibnamefont{Thompson}},
  \bibnamefont{and} \bibinfo{author}{\bibfnamefont{Q.}~\bibnamefont{Si}},
  \bibinfo{journal}{Rev.\ Mod.\ Phys.} \textbf{\bibinfo{volume}{92}},
  \bibinfo{pages}{011002} (\bibinfo{year}{2020}).

\bibitem[{\citenamefont{Wirth et~al.}(2012)\citenamefont{Wirth, Ernst,
  Cardoso-Gil, Borrmann, Seiro, Krellner, Geibel, Kirchner, Burkhardt, Grin
  et~al.}}]{wir12}
\bibinfo{author}{\bibfnamefont{S.}~\bibnamefont{Wirth}},
  \bibinfo{author}{\bibfnamefont{S.}~\bibnamefont{Ernst}},
  \bibinfo{author}{\bibfnamefont{R.}~\bibnamefont{Cardoso-Gil}},
  \bibinfo{author}{\bibfnamefont{H.}~\bibnamefont{Borrmann}},
  \bibinfo{author}{\bibfnamefont{S.}~\bibnamefont{Seiro}},
  \bibinfo{author}{\bibfnamefont{C.}~\bibnamefont{Krellner}},
  \bibinfo{author}{\bibfnamefont{C.}~\bibnamefont{Geibel}},
  \bibinfo{author}{\bibfnamefont{S.}~\bibnamefont{Kirchner}},
  \bibinfo{author}{\bibfnamefont{U.}~\bibnamefont{Burkhardt}},
  \bibinfo{author}{\bibfnamefont{Y.}~\bibnamefont{Grin}}, \bibnamefont{et~al.},
  \bibinfo{journal}{J.\ Phys.: Condens.\ Matter} \textbf{\bibinfo{volume}{24}},
  \bibinfo{pages}{294203} (\bibinfo{year}{2012}).

\bibitem[{\citenamefont{Alexandrov et~al.}(2015)\citenamefont{Alexandrov,
  Coleman, and Erten}}]{ale15}
\bibinfo{author}{\bibfnamefont{V.}~\bibnamefont{Alexandrov}},
  \bibinfo{author}{\bibfnamefont{P.}~\bibnamefont{Coleman}}, \bibnamefont{and}
  \bibinfo{author}{\bibfnamefont{O.}~\bibnamefont{Erten}},
  \bibinfo{journal}{Phys.\ Rev.\ Lett.} \textbf{\bibinfo{volume}{114}},
  \bibinfo{pages}{177202} (\bibinfo{year}{2015}).

\bibitem[{\citenamefont{Stockert
  et~al.}(2006{\natexlab{a}})\citenamefont{Stockert, Koza, Ferstl, Murani,
  Geibel, and Steglich}}]{sto06a}
\bibinfo{author}{\bibfnamefont{O.}~\bibnamefont{Stockert}},
  \bibinfo{author}{\bibfnamefont{M.~M.} \bibnamefont{Koza}},
  \bibinfo{author}{\bibfnamefont{J.}~\bibnamefont{Ferstl}},
  \bibinfo{author}{\bibfnamefont{A.~P.} \bibnamefont{Murani}},
  \bibinfo{author}{\bibfnamefont{C.}~\bibnamefont{Geibel}}, \bibnamefont{and}
  \bibinfo{author}{\bibfnamefont{F.}~\bibnamefont{Steglich}},
  \bibinfo{journal}{Physica B} \textbf{\bibinfo{volume}{378}},
  \bibinfo{pages}{157} (\bibinfo{year}{2006}{\natexlab{a}}).

\bibitem[{\citenamefont{Zwicknagl}(2011)}]{zwi11}
\bibinfo{author}{\bibfnamefont{G.}~\bibnamefont{Zwicknagl}},
  \bibinfo{journal}{J. Phys.\ Condens. Matter} \textbf{\bibinfo{volume}{23}},
  \bibinfo{pages}{094215} (\bibinfo{year}{2011}).

\bibitem[{\citenamefont{Coleman et~al.}(1985)\citenamefont{Coleman, Anderson,
  and Ramakrishnan}}]{col85}
\bibinfo{author}{\bibfnamefont{P.}~\bibnamefont{Coleman}},
  \bibinfo{author}{\bibfnamefont{P.~W.} \bibnamefont{Anderson}},
  \bibnamefont{and} \bibinfo{author}{\bibfnamefont{T.~V.}
  \bibnamefont{Ramakrishnan}}, \bibinfo{journal}{Phys.\ Rev.\ Lett.}
  \textbf{\bibinfo{volume}{55}}, \bibinfo{pages}{414} (\bibinfo{year}{1985}).

\bibitem[{\citenamefont{Sun and Steglich}(2013)}]{sun13}
\bibinfo{author}{\bibfnamefont{P.}~\bibnamefont{Sun}} \bibnamefont{and}
  \bibinfo{author}{\bibfnamefont{F.}~\bibnamefont{Steglich}},
  \bibinfo{journal}{Phys.\ Rev.\ Lett.} \textbf{\bibinfo{volume}{110}},
  \bibinfo{pages}{216408} (\bibinfo{year}{2013}).

\bibitem[{\citenamefont{Hartmann et~al.}(2010)\citenamefont{Hartmann, Oeschler,
  Krellner, Geibel, Paschen, and Steglich}}]{har10}
\bibinfo{author}{\bibfnamefont{S.}~\bibnamefont{Hartmann}},
  \bibinfo{author}{\bibfnamefont{N.}~\bibnamefont{Oeschler}},
  \bibinfo{author}{\bibfnamefont{C.}~\bibnamefont{Krellner}},
  \bibinfo{author}{\bibfnamefont{C.}~\bibnamefont{Geibel}},
  \bibinfo{author}{\bibfnamefont{S.}~\bibnamefont{Paschen}}, \bibnamefont{and}
  \bibinfo{author}{\bibfnamefont{F.}~\bibnamefont{Steglich}},
  \bibinfo{journal}{Phys.\ Rev.\ Lett.} \textbf{\bibinfo{volume}{104}},
  \bibinfo{pages}{096401} (\bibinfo{year}{2010}).

\bibitem[{\citenamefont{Chen et~al.}(2018)\citenamefont{Chen, Xu, Niu, Peng,
  Xu, Wen, Liu, Shu, Tan, Lai et~al.}}]{che18}
\bibinfo{author}{\bibfnamefont{Q.~Y.} \bibnamefont{Chen}},
  \bibinfo{author}{\bibfnamefont{D.~F.} \bibnamefont{Xu}},
  \bibinfo{author}{\bibfnamefont{X.~H.} \bibnamefont{Niu}},
  \bibinfo{author}{\bibfnamefont{R.}~\bibnamefont{Peng}},
  \bibinfo{author}{\bibfnamefont{H.~C.} \bibnamefont{Xu}},
  \bibinfo{author}{\bibfnamefont{C.~H.~P.} \bibnamefont{Wen}},
  \bibinfo{author}{\bibfnamefont{X.}~\bibnamefont{Liu}},
  \bibinfo{author}{\bibfnamefont{L.}~\bibnamefont{Shu}},
  \bibinfo{author}{\bibfnamefont{S.~Y.} \bibnamefont{Tan}},
  \bibinfo{author}{\bibfnamefont{X.~C.} \bibnamefont{Lai}},
  \bibnamefont{et~al.}, \bibinfo{journal}{Phys.\ Rev.\ Lett.}
  \textbf{\bibinfo{volume}{120}}, \bibinfo{pages}{066403}
  (\bibinfo{year}{2018}).

\bibitem[{\citenamefont{Pfau et~al.}(2012)\citenamefont{Pfau, Hartmann,
  Stockert, Sun, Lausberg, Brando, Friedemann, Krellner, Geibel, Wirth
  et~al.}}]{pfau12}
\bibinfo{author}{\bibfnamefont{H.}~\bibnamefont{Pfau}},
  \bibinfo{author}{\bibfnamefont{S.}~\bibnamefont{Hartmann}},
  \bibinfo{author}{\bibfnamefont{U.}~\bibnamefont{Stockert}},
  \bibinfo{author}{\bibfnamefont{P.}~\bibnamefont{Sun}},
  \bibinfo{author}{\bibfnamefont{S.}~\bibnamefont{Lausberg}},
  \bibinfo{author}{\bibfnamefont{M.}~\bibnamefont{Brando}},
  \bibinfo{author}{\bibfnamefont{S.}~\bibnamefont{Friedemann}},
  \bibinfo{author}{\bibfnamefont{C.}~\bibnamefont{Krellner}},
  \bibinfo{author}{\bibfnamefont{C.}~\bibnamefont{Geibel}},
  \bibinfo{author}{\bibfnamefont{S.}~\bibnamefont{Wirth}},
  \bibnamefont{et~al.}, \bibinfo{journal}{Nature}
  \textbf{\bibinfo{volume}{484}}, \bibinfo{pages}{493} (\bibinfo{year}{2012}).

\bibitem[{\citenamefont{Pourret et~al.}(2014)\citenamefont{Pourret, Aoki,
  Boukahil, Brison, Knafo, Knebel, Raymond, Taupin, {\=O}nuki, and
  Flouquet}}]{pou14}
\bibinfo{author}{\bibfnamefont{A.}~\bibnamefont{Pourret}},
  \bibinfo{author}{\bibfnamefont{D.}~\bibnamefont{Aoki}},
  \bibinfo{author}{\bibfnamefont{M.}~\bibnamefont{Boukahil}},
  \bibinfo{author}{\bibfnamefont{J.-P.} \bibnamefont{Brison}},
  \bibinfo{author}{\bibfnamefont{W.}~\bibnamefont{Knafo}},
  \bibinfo{author}{\bibfnamefont{G.}~\bibnamefont{Knebel}},
  \bibinfo{author}{\bibfnamefont{S.}~\bibnamefont{Raymond}},
  \bibinfo{author}{\bibfnamefont{M.}~\bibnamefont{Taupin}},
  \bibinfo{author}{\bibfnamefont{Y.}~\bibnamefont{{\=O}nuki}},
  \bibnamefont{and} \bibinfo{author}{\bibfnamefont{J.}~\bibnamefont{Flouquet}},
  \bibinfo{journal}{J.\ Phys.\ Soc.\ Jpn.} \textbf{\bibinfo{volume}{83}},
  \bibinfo{pages}{061002} (\bibinfo{year}{2014}).

\bibitem[{\citenamefont{Parfen’eva et~al.}(2004)\citenamefont{Parfen’eva,
  Smirnov, Misiorek, Mucha, Jezowski, Prokof’ev, and Assmus}}]{par04}
\bibinfo{author}{\bibfnamefont{L.~S.} \bibnamefont{Parfen’eva}},
  \bibinfo{author}{\bibfnamefont{I.~A.} \bibnamefont{Smirnov}},
  \bibinfo{author}{\bibfnamefont{H.}~\bibnamefont{Misiorek}},
  \bibinfo{author}{\bibfnamefont{J.}~\bibnamefont{Mucha}},
  \bibinfo{author}{\bibfnamefont{A.}~\bibnamefont{Jezowski}},
  \bibinfo{author}{\bibfnamefont{A.~V.} \bibnamefont{Prokof’ev}},
  \bibnamefont{and} \bibinfo{author}{\bibfnamefont{W.}~\bibnamefont{Assmus}},
  \bibinfo{journal}{Phys.\ Solid State} \textbf{\bibinfo{volume}{46}},
  \bibinfo{pages}{357} (\bibinfo{year}{2004}).

\bibitem[{\citenamefont{Singh et~al.}(2007)\citenamefont{Singh, Capan, Nicklas,
  Rams, Gladun, Lee, DiTusa, Fisk, Steglich, and Wirth}}]{sin07}
\bibinfo{author}{\bibfnamefont{S.}~\bibnamefont{Singh}},
  \bibinfo{author}{\bibfnamefont{C.}~\bibnamefont{Capan}},
  \bibinfo{author}{\bibfnamefont{M.}~\bibnamefont{Nicklas}},
  \bibinfo{author}{\bibfnamefont{M.}~\bibnamefont{Rams}},
  \bibinfo{author}{\bibfnamefont{A.}~\bibnamefont{Gladun}},
  \bibinfo{author}{\bibfnamefont{H.-O.} \bibnamefont{Lee}},
  \bibinfo{author}{\bibfnamefont{J.~F.} \bibnamefont{DiTusa}},
  \bibinfo{author}{\bibfnamefont{Z.}~\bibnamefont{Fisk}},
  \bibinfo{author}{\bibfnamefont{F.}~\bibnamefont{Steglich}}, \bibnamefont{and}
  \bibinfo{author}{\bibfnamefont{S.}~\bibnamefont{Wirth}},
  \bibinfo{journal}{Phys.\ Rev.\ Lett.} \textbf{\bibinfo{volume}{98}},
  \bibinfo{pages}{057001} (\bibinfo{year}{2007}).

\bibitem[{\citenamefont{Zaum et~al.}(2011)\citenamefont{Zaum, Grube,
  Sch{\"a}fer, Bauer, Thompson, and {von L{\"o}hneysen}}}]{zau11}
\bibinfo{author}{\bibfnamefont{S.}~\bibnamefont{Zaum}},
  \bibinfo{author}{\bibfnamefont{K.}~\bibnamefont{Grube}},
  \bibinfo{author}{\bibfnamefont{R.}~\bibnamefont{Sch{\"a}fer}},
  \bibinfo{author}{\bibfnamefont{E.~D.} \bibnamefont{Bauer}},
  \bibinfo{author}{\bibfnamefont{J.~D.} \bibnamefont{Thompson}},
  \bibnamefont{and} \bibinfo{author}{\bibfnamefont{H.}~\bibnamefont{{von
  L{\"o}hneysen}}}, \bibinfo{journal}{Phys.\ Rev.\ Lett.}
  \textbf{\bibinfo{volume}{106}}, \bibinfo{pages}{087003}
  (\bibinfo{year}{2011}).

\bibitem[{\citenamefont{Tanatar et~al.}(2007)\citenamefont{Tanatar, Paglione,
  Petrovic, and Taillefer}}]{tan07}
\bibinfo{author}{\bibfnamefont{M.~A.} \bibnamefont{Tanatar}},
  \bibinfo{author}{\bibfnamefont{J.}~\bibnamefont{Paglione}},
  \bibinfo{author}{\bibfnamefont{C.}~\bibnamefont{Petrovic}}, \bibnamefont{and}
  \bibinfo{author}{\bibfnamefont{L.}~\bibnamefont{Taillefer}},
  \bibinfo{journal}{Science} \textbf{\bibinfo{volume}{316}},
  \bibinfo{pages}{1320} (\bibinfo{year}{2007}).

\bibitem[{\citenamefont{Smith and McKenzie}(2008)}]{smi08}
\bibinfo{author}{\bibfnamefont{M.~F.} \bibnamefont{Smith}} \bibnamefont{and}
  \bibinfo{author}{\bibfnamefont{R.~H.} \bibnamefont{McKenzie}},
  \bibinfo{journal}{Phys.\ Rev.\ Lett.} \textbf{\bibinfo{volume}{101}},
  \bibinfo{pages}{266403} (\bibinfo{year}{2008}).

\bibitem[{\citenamefont{Smidman et~al.}(2018)\citenamefont{Smidman, Stockert,
  Arndt, Pang, Jiao, Yuan, Vieyra, Kitagawa, Ishida, Fujiwara et~al.}}]{smi18}
\bibinfo{author}{\bibfnamefont{M.}~\bibnamefont{Smidman}},
  \bibinfo{author}{\bibfnamefont{O.}~\bibnamefont{Stockert}},
  \bibinfo{author}{\bibfnamefont{J.}~\bibnamefont{Arndt}},
  \bibinfo{author}{\bibfnamefont{G.~M.} \bibnamefont{Pang}},
  \bibinfo{author}{\bibfnamefont{L.}~\bibnamefont{Jiao}},
  \bibinfo{author}{\bibfnamefont{H.~Q.} \bibnamefont{Yuan}},
  \bibinfo{author}{\bibfnamefont{H.~A.} \bibnamefont{Vieyra}},
  \bibinfo{author}{\bibfnamefont{S.}~\bibnamefont{Kitagawa}},
  \bibinfo{author}{\bibfnamefont{K.}~\bibnamefont{Ishida}},
  \bibinfo{author}{\bibfnamefont{K.}~\bibnamefont{Fujiwara}},
  \bibnamefont{et~al.}, \bibinfo{journal}{Phil.\ Mag.}
  \textbf{\bibinfo{volume}{98}}, \bibinfo{pages}{2930} (\bibinfo{year}{2018}).

\bibitem[{\citenamefont{Steglich et~al.}(2014)\citenamefont{Steglich, Pfau,
  Lausberg, Hamann, Sun, Stockert, Brando, Friedemann, Krellner, Geibel
  et~al.}}]{steg14}
\bibinfo{author}{\bibfnamefont{F.}~\bibnamefont{Steglich}},
  \bibinfo{author}{\bibfnamefont{H.}~\bibnamefont{Pfau}},
  \bibinfo{author}{\bibfnamefont{S.}~\bibnamefont{Lausberg}},
  \bibinfo{author}{\bibfnamefont{S.}~\bibnamefont{Hamann}},
  \bibinfo{author}{\bibfnamefont{P.}~\bibnamefont{Sun}},
  \bibinfo{author}{\bibfnamefont{U.}~\bibnamefont{Stockert}},
  \bibinfo{author}{\bibfnamefont{M.}~\bibnamefont{Brando}},
  \bibinfo{author}{\bibfnamefont{S.}~\bibnamefont{Friedemann}},
  \bibinfo{author}{\bibfnamefont{C.}~\bibnamefont{Krellner}},
  \bibinfo{author}{\bibfnamefont{C.}~\bibnamefont{Geibel}},
  \bibnamefont{et~al.}, \bibinfo{journal}{J.\ Phys.\ Soc.\ Jpn.}
  \textbf{\bibinfo{volume}{83}}, \bibinfo{pages}{061001}
  (\bibinfo{year}{2014}).

\bibitem[{\citenamefont{Machida et~al.}(2013)\citenamefont{Machida, Tomokuni,
  Izawa, Lapertot, Knebel, Brison, and Flouquet}}]{mac13}
\bibinfo{author}{\bibfnamefont{Y.}~\bibnamefont{Machida}},
  \bibinfo{author}{\bibfnamefont{K.}~\bibnamefont{Tomokuni}},
  \bibinfo{author}{\bibfnamefont{K.}~\bibnamefont{Izawa}},
  \bibinfo{author}{\bibfnamefont{G.}~\bibnamefont{Lapertot}},
  \bibinfo{author}{\bibfnamefont{G.}~\bibnamefont{Knebel}},
  \bibinfo{author}{\bibfnamefont{J.-P.} \bibnamefont{Brison}},
  \bibnamefont{and} \bibinfo{author}{\bibfnamefont{J.}~\bibnamefont{Flouquet}},
  \bibinfo{journal}{Phys.\ Rev.\ Lett.} \textbf{\bibinfo{volume}{110}},
  \bibinfo{pages}{236402} (\bibinfo{year}{2013}).

\bibitem[{\citenamefont{Reid et~al.}(2014)\citenamefont{Reid, Tanatar, Daou,
  Hu, Petrovic, and Taillefer}}]{rei14}
\bibinfo{author}{\bibfnamefont{J.-P.} \bibnamefont{Reid}},
  \bibinfo{author}{\bibfnamefont{M.~A.} \bibnamefont{Tanatar}},
  \bibinfo{author}{\bibfnamefont{R.}~\bibnamefont{Daou}},
  \bibinfo{author}{\bibfnamefont{R.}~\bibnamefont{Hu}},
  \bibinfo{author}{\bibfnamefont{C.}~\bibnamefont{Petrovic}}, \bibnamefont{and}
  \bibinfo{author}{\bibfnamefont{L.}~\bibnamefont{Taillefer}},
  \bibinfo{journal}{Phys.\ Rev.\ B} \textbf{\bibinfo{volume}{89}},
  \bibinfo{pages}{045130} (\bibinfo{year}{2014}).

\bibitem[{\citenamefont{Taupin et~al.}(2015)\citenamefont{Taupin, Knebel,
  Matsuda, Lapertot, Machida, Izawa, Brison, and Flouquet}}]{tau15}
\bibinfo{author}{\bibfnamefont{M.}~\bibnamefont{Taupin}},
  \bibinfo{author}{\bibfnamefont{G.}~\bibnamefont{Knebel}},
  \bibinfo{author}{\bibfnamefont{T.~D.} \bibnamefont{Matsuda}},
  \bibinfo{author}{\bibfnamefont{G.}~\bibnamefont{Lapertot}},
  \bibinfo{author}{\bibfnamefont{Y.}~\bibnamefont{Machida}},
  \bibinfo{author}{\bibfnamefont{K.}~\bibnamefont{Izawa}},
  \bibinfo{author}{\bibfnamefont{J.-P.} \bibnamefont{Brison}},
  \bibnamefont{and} \bibinfo{author}{\bibfnamefont{J.}~\bibnamefont{Flouquet}},
  \bibinfo{journal}{Phys.\ Rev.\ Lett.} \textbf{\bibinfo{volume}{115}},
  \bibinfo{pages}{046402} (\bibinfo{year}{2015}).

\bibitem[{\citenamefont{Dong et~al.}(2013)\citenamefont{Dong, Tokiwa, Bud'ko,
  Canfield, and Gegenwart}}]{dong13}
\bibinfo{author}{\bibfnamefont{J.~K.} \bibnamefont{Dong}},
  \bibinfo{author}{\bibfnamefont{Y.}~\bibnamefont{Tokiwa}},
  \bibinfo{author}{\bibfnamefont{S.~L.} \bibnamefont{Bud'ko}},
  \bibinfo{author}{\bibfnamefont{P.~C.} \bibnamefont{Canfield}},
  \bibnamefont{and}
  \bibinfo{author}{\bibfnamefont{P.}~\bibnamefont{Gegenwart}},
  \bibinfo{journal}{Phys.\ Rev.\ Lett.} \textbf{\bibinfo{volume}{110}},
  \bibinfo{pages}{176402} (\bibinfo{year}{2013}).

\bibitem[{\citenamefont{Lausberg}(2013)}]{lau13}
\bibinfo{author}{\bibfnamefont{S.}~\bibnamefont{Lausberg}}, Ph.D. thesis,
  \bibinfo{school}{Technical University Dresden, Germany}
  (\bibinfo{year}{2013}).

\bibitem[{\citenamefont{Steglich et~al.}(1979)\citenamefont{Steglich, Aarts,
  Bredl, Lieke, Meschede, Franz, and Sch{\"a}fer}}]{ste79}
\bibinfo{author}{\bibfnamefont{F.}~\bibnamefont{Steglich}},
  \bibinfo{author}{\bibfnamefont{J.}~\bibnamefont{Aarts}},
  \bibinfo{author}{\bibfnamefont{C.}~\bibnamefont{Bredl}},
  \bibinfo{author}{\bibfnamefont{W.}~\bibnamefont{Lieke}},
  \bibinfo{author}{\bibfnamefont{D.}~\bibnamefont{Meschede}},
  \bibinfo{author}{\bibfnamefont{W.}~\bibnamefont{Franz}}, \bibnamefont{and}
  \bibinfo{author}{\bibfnamefont{H.}~\bibnamefont{Sch{\"a}fer}},
  \bibinfo{journal}{Phys.\ Rev.\ Lett.} \textbf{\bibinfo{volume}{43}},
  \bibinfo{pages}{1892} (\bibinfo{year}{1979}).

\bibitem[{\citenamefont{Rauchschwalbe et~al.}(1982)\citenamefont{Rauchschwalbe,
  Lieke, Bredl, Steglich, Aarts, Martini, and Mota}}]{rau82}
\bibinfo{author}{\bibfnamefont{U.}~\bibnamefont{Rauchschwalbe}},
  \bibinfo{author}{\bibfnamefont{W.}~\bibnamefont{Lieke}},
  \bibinfo{author}{\bibfnamefont{C.~D.} \bibnamefont{Bredl}},
  \bibinfo{author}{\bibfnamefont{F.}~\bibnamefont{Steglich}},
  \bibinfo{author}{\bibfnamefont{J.}~\bibnamefont{Aarts}},
  \bibinfo{author}{\bibfnamefont{K.~M.} \bibnamefont{Martini}},
  \bibnamefont{and} \bibinfo{author}{\bibfnamefont{A.~C.} \bibnamefont{Mota}},
  \bibinfo{journal}{Phys.\ Rev.\ Lett.} \textbf{\bibinfo{volume}{49}},
  \bibinfo{pages}{1448} (\bibinfo{year}{1982}).

\bibitem[{\citenamefont{Assmus et~al.}(1984)\citenamefont{Assmus, Herrmann,
  Rauchschwalbe, Riegel, Lieke, Spille, Horn, Weber, Steglich, and
  Cordier}}]{ass84}
\bibinfo{author}{\bibfnamefont{W.}~\bibnamefont{Assmus}},
  \bibinfo{author}{\bibfnamefont{M.}~\bibnamefont{Herrmann}},
  \bibinfo{author}{\bibfnamefont{U.}~\bibnamefont{Rauchschwalbe}},
  \bibinfo{author}{\bibfnamefont{S.}~\bibnamefont{Riegel}},
  \bibinfo{author}{\bibfnamefont{W.}~\bibnamefont{Lieke}},
  \bibinfo{author}{\bibfnamefont{H.}~\bibnamefont{Spille}},
  \bibinfo{author}{\bibfnamefont{S.}~\bibnamefont{Horn}},
  \bibinfo{author}{\bibfnamefont{G.}~\bibnamefont{Weber}},
  \bibinfo{author}{\bibfnamefont{F.}~\bibnamefont{Steglich}}, \bibnamefont{and}
  \bibinfo{author}{\bibfnamefont{G.}~\bibnamefont{Cordier}},
  \bibinfo{journal}{Phys.\ Rev.\ Lett.} \textbf{\bibinfo{volume}{52}},
  \bibinfo{pages}{469} (\bibinfo{year}{1984}).

\bibitem[{\citenamefont{Coleman}(2015)}]{col15b}
\bibinfo{author}{\bibfnamefont{P.}~\bibnamefont{Coleman}}
  (\bibinfo{year}{2015}), \bibinfo{note}{{p}rivate communication}.

\bibitem[{\citenamefont{Sichelschmidt et~al.}(2003)\citenamefont{Sichelschmidt,
  Ivanshin, Ferstl, Geibel, and Steglich}}]{sic03}
\bibinfo{author}{\bibfnamefont{J.}~\bibnamefont{Sichelschmidt}},
  \bibinfo{author}{\bibfnamefont{V.~A.} \bibnamefont{Ivanshin}},
  \bibinfo{author}{\bibfnamefont{J.}~\bibnamefont{Ferstl}},
  \bibinfo{author}{\bibfnamefont{C.}~\bibnamefont{Geibel}}, \bibnamefont{and}
  \bibinfo{author}{\bibfnamefont{F.}~\bibnamefont{Steglich}},
  \bibinfo{journal}{Phys.\ Rev.\ Lett.} \textbf{\bibinfo{volume}{91}},
  \bibinfo{pages}{156401} (\bibinfo{year}{2003}).

\bibitem[{\citenamefont{Hamann et~al.}(2019)\citenamefont{Hamann, Zhang, Jang,
  Hannaske, Steinke, Lausberg, Pedrero, Klingner, Baenitz, Steglich
  et~al.}}]{ham19}
\bibinfo{author}{\bibfnamefont{S.}~\bibnamefont{Hamann}},
  \bibinfo{author}{\bibfnamefont{J.}~\bibnamefont{Zhang}},
  \bibinfo{author}{\bibfnamefont{D.}~\bibnamefont{Jang}},
  \bibinfo{author}{\bibfnamefont{A.}~\bibnamefont{Hannaske}},
  \bibinfo{author}{\bibfnamefont{L.}~\bibnamefont{Steinke}},
  \bibinfo{author}{\bibfnamefont{S.}~\bibnamefont{Lausberg}},
  \bibinfo{author}{\bibfnamefont{L.}~\bibnamefont{Pedrero}},
  \bibinfo{author}{\bibfnamefont{C.}~\bibnamefont{Klingner}},
  \bibinfo{author}{\bibfnamefont{M.}~\bibnamefont{Baenitz}},
  \bibinfo{author}{\bibfnamefont{F.}~\bibnamefont{Steglich}},
  \bibnamefont{et~al.}, \bibinfo{journal}{Phys.\ Rev.\ Lett.}
  \textbf{\bibinfo{volume}{122}}, \bibinfo{pages}{077202}
  (\bibinfo{year}{2019}).

\bibitem[{\citenamefont{Saunders}(2018)}]{sau18}
\bibinfo{author}{\bibfnamefont{J.}~\bibnamefont{Saunders}}
  (\bibinfo{year}{2018}), \bibinfo{note}{invited Talk at 12th Intern.\ Conf.\
  on Materials and Mechanisms of Superconductivity (M$^2$S 2018), Beijing,
  China}.

\bibitem[{\citenamefont{Knebel et~al.}(2006)\citenamefont{Knebel, Boursier,
  Hassinger, Lapertot, Niklowitz, Pourret, Salce, Sanchez, Sheikin, Bonville
  et~al.}}]{kne06}
\bibinfo{author}{\bibfnamefont{G.}~\bibnamefont{Knebel}},
  \bibinfo{author}{\bibfnamefont{R.}~\bibnamefont{Boursier}},
  \bibinfo{author}{\bibfnamefont{E.}~\bibnamefont{Hassinger}},
  \bibinfo{author}{\bibfnamefont{G.}~\bibnamefont{Lapertot}},
  \bibinfo{author}{\bibfnamefont{P.~G.} \bibnamefont{Niklowitz}},
  \bibinfo{author}{\bibfnamefont{A.}~\bibnamefont{Pourret}},
  \bibinfo{author}{\bibfnamefont{B.}~\bibnamefont{Salce}},
  \bibinfo{author}{\bibfnamefont{J.~P.} \bibnamefont{Sanchez}},
  \bibinfo{author}{\bibfnamefont{I.}~\bibnamefont{Sheikin}},
  \bibinfo{author}{\bibfnamefont{P.}~\bibnamefont{Bonville}},
  \bibnamefont{et~al.}, \bibinfo{journal}{J.\ Phys.\ Soc.\ Jpn.}
  \textbf{\bibinfo{volume}{75}}, \bibinfo{pages}{114709}
  (\bibinfo{year}{2006}).

\bibitem[{\citenamefont{Park et~al.}(2012)\citenamefont{Park, Lee, Martin, Lu,
  Sidorov, Gofryk, Ronning, Bauer, and Thompson}}]{par12}
\bibinfo{author}{\bibfnamefont{T.}~\bibnamefont{Park}},
  \bibinfo{author}{\bibfnamefont{H.}~\bibnamefont{Lee}},
  \bibinfo{author}{\bibfnamefont{I.}~\bibnamefont{Martin}},
  \bibinfo{author}{\bibfnamefont{X.}~\bibnamefont{Lu}},
  \bibinfo{author}{\bibfnamefont{V.~A.} \bibnamefont{Sidorov}},
  \bibinfo{author}{\bibfnamefont{K.}~\bibnamefont{Gofryk}},
  \bibinfo{author}{\bibfnamefont{F.}~\bibnamefont{Ronning}},
  \bibinfo{author}{\bibfnamefont{E.~D.} \bibnamefont{Bauer}}, \bibnamefont{and}
  \bibinfo{author}{\bibfnamefont{J.~D.} \bibnamefont{Thompson}},
  \bibinfo{journal}{Phys.\ Rev.\ Lett.} \textbf{\bibinfo{volume}{108}},
  \bibinfo{pages}{077003} (\bibinfo{year}{2012}).

\bibitem[{\citenamefont{Petrovic et~al.}(2001)\citenamefont{Petrovic,
  Movshovich, Jaime, Pagliuso, Hundley, Sarrao, Fisk, and Thompson}}]{pet01}
\bibinfo{author}{\bibfnamefont{C.}~\bibnamefont{Petrovic}},
  \bibinfo{author}{\bibfnamefont{R.}~\bibnamefont{Movshovich}},
  \bibinfo{author}{\bibfnamefont{M.}~\bibnamefont{Jaime}},
  \bibinfo{author}{\bibfnamefont{P.~G.} \bibnamefont{Pagliuso}},
  \bibinfo{author}{\bibfnamefont{M.~F.} \bibnamefont{Hundley}},
  \bibinfo{author}{\bibfnamefont{J.~L.} \bibnamefont{Sarrao}},
  \bibinfo{author}{\bibfnamefont{Z.}~\bibnamefont{Fisk}}, \bibnamefont{and}
  \bibinfo{author}{\bibfnamefont{J.~D.} \bibnamefont{Thompson}},
  \bibinfo{journal}{Europhys.\ Lett.} \textbf{\bibinfo{volume}{53}},
  \bibinfo{pages}{354} (\bibinfo{year}{2001}).

\bibitem[{\citenamefont{Gegenwart et~al.}(2005)\citenamefont{Gegenwart,
  Custers, Tokiwa, Geibel, and Steglich}}]{geg05}
\bibinfo{author}{\bibfnamefont{P.}~\bibnamefont{Gegenwart}},
  \bibinfo{author}{\bibfnamefont{J.}~\bibnamefont{Custers}},
  \bibinfo{author}{\bibfnamefont{Y.}~\bibnamefont{Tokiwa}},
  \bibinfo{author}{\bibfnamefont{C.}~\bibnamefont{Geibel}}, \bibnamefont{and}
  \bibinfo{author}{\bibfnamefont{F.}~\bibnamefont{Steglich}},
  \bibinfo{journal}{Phys.\ Rev.\ Lett.} \textbf{\bibinfo{volume}{94}},
  \bibinfo{pages}{076402} (\bibinfo{year}{2005}).

\bibitem[{\citenamefont{Stockert
  et~al.}(2006{\natexlab{b}})\citenamefont{Stockert, Andreica, Amato, Jeevan,
  Geibel, and Steglich}}]{sto06}
\bibinfo{author}{\bibfnamefont{O.}~\bibnamefont{Stockert}},
  \bibinfo{author}{\bibfnamefont{D.}~\bibnamefont{Andreica}},
  \bibinfo{author}{\bibfnamefont{A.}~\bibnamefont{Amato}},
  \bibinfo{author}{\bibfnamefont{H.~S.} \bibnamefont{Jeevan}},
  \bibinfo{author}{\bibfnamefont{C.}~\bibnamefont{Geibel}}, \bibnamefont{and}
  \bibinfo{author}{\bibfnamefont{F.}~\bibnamefont{Steglich}},
  \bibinfo{journal}{Physica B} \textbf{\bibinfo{volume}{374-375}},
  \bibinfo{pages}{167} (\bibinfo{year}{2006}{\natexlab{b}}).

\bibitem[{\citenamefont{Li et~al.}(2019)\citenamefont{Li, Wang, Xu, Xie, and
  f.~Yang}}]{li19}
\bibinfo{author}{\bibfnamefont{Y.}~\bibnamefont{Li}},
  \bibinfo{author}{\bibfnamefont{Q.}~\bibnamefont{Wang}},
  \bibinfo{author}{\bibfnamefont{Y.}~\bibnamefont{Xu}},
  \bibinfo{author}{\bibfnamefont{W.}~\bibnamefont{Xie}}, \bibnamefont{and}
  \bibinfo{author}{\bibfnamefont{Y.}~\bibnamefont{f.~Yang}},
  \bibinfo{journal}{Phys.\ Rev.\ B} \textbf{\bibinfo{volume}{100}},
  \bibinfo{pages}{085132} (\bibinfo{year}{2019}).

\bibitem[{\citenamefont{Hu et~al.}(2021{\natexlab{a}})\citenamefont{Hu, Cai,
  Chen, and Si}}]{hu21}
\bibinfo{author}{\bibfnamefont{H.}~\bibnamefont{Hu}},
  \bibinfo{author}{\bibfnamefont{A.}~\bibnamefont{Cai}},
  \bibinfo{author}{\bibfnamefont{L.}~\bibnamefont{Chen}}, \bibnamefont{and}
  \bibinfo{author}{\bibfnamefont{Q.}~\bibnamefont{Si}}
  (\bibinfo{year}{2021}{\natexlab{a}}),
  \bibinfo{note}{\textit{{a}rXiv}:2109.12794}.

\bibitem[{\citenamefont{Sarrao et~al.}(2002)\citenamefont{Sarrao, Morales,
  Thompson, Scott, Stewart, Wastin, Rebizant, Boulet, Colineau, and
  Lander}}]{sar02}
\bibinfo{author}{\bibfnamefont{J.~L.} \bibnamefont{Sarrao}},
  \bibinfo{author}{\bibfnamefont{L.~A.} \bibnamefont{Morales}},
  \bibinfo{author}{\bibfnamefont{J.~D.} \bibnamefont{Thompson}},
  \bibinfo{author}{\bibfnamefont{B.~L.} \bibnamefont{Scott}},
  \bibinfo{author}{\bibfnamefont{G.~R.} \bibnamefont{Stewart}},
  \bibinfo{author}{\bibfnamefont{F.}~\bibnamefont{Wastin}},
  \bibinfo{author}{\bibfnamefont{J.}~\bibnamefont{Rebizant}},
  \bibinfo{author}{\bibfnamefont{P.}~\bibnamefont{Boulet}},
  \bibinfo{author}{\bibfnamefont{E.}~\bibnamefont{Colineau}}, \bibnamefont{and}
  \bibinfo{author}{\bibfnamefont{G.~H.} \bibnamefont{Lander}},
  \bibinfo{journal}{Nature} \textbf{\bibinfo{volume}{420}},
  \bibinfo{pages}{297} (\bibinfo{year}{2002}).

\bibitem[{\citenamefont{Hu et~al.}(2021{\natexlab{b}})\citenamefont{Hu, Cai,
  Chen, Deng, Pixley, Ingersent, and Si}}]{hu21b}
\bibinfo{author}{\bibfnamefont{H.}~\bibnamefont{Hu}},
  \bibinfo{author}{\bibfnamefont{A.}~\bibnamefont{Cai}},
  \bibinfo{author}{\bibfnamefont{L.}~\bibnamefont{Chen}},
  \bibinfo{author}{\bibfnamefont{L.}~\bibnamefont{Deng}},
  \bibinfo{author}{\bibfnamefont{J.~H.} \bibnamefont{Pixley}},
  \bibinfo{author}{\bibfnamefont{K.}~\bibnamefont{Ingersent}},
  \bibnamefont{and} \bibinfo{author}{\bibfnamefont{Q.}~\bibnamefont{Si}}
  (\bibinfo{year}{2021}{\natexlab{b}}),
  \bibinfo{note}{\textit{{a}rXiv}:2109.13224}.

\bibitem[{\citenamefont{Maple et~al.}(2004)\citenamefont{Maple, Bauer, Zapf,
  and Wosnitza}}]{map04}
\bibinfo{author}{\bibfnamefont{M.~B.} \bibnamefont{Maple}},
  \bibinfo{author}{\bibfnamefont{E.~D.} \bibnamefont{Bauer}},
  \bibinfo{author}{\bibfnamefont{V.~S.} \bibnamefont{Zapf}}, \bibnamefont{and}
  \bibinfo{author}{\bibfnamefont{J.}~\bibnamefont{Wosnitza}}, in
  \emph{\bibinfo{booktitle}{The Physics of Superconductors}}, edited by
  \bibinfo{editor}{\bibfnamefont{K.~H.} \bibnamefont{Bennemann}}
  \bibnamefont{and} \bibinfo{editor}{\bibfnamefont{J.~B.}
  \bibnamefont{Ketterson}} (\bibinfo{publisher}{Springer},
  \bibinfo{address}{Berlin, Heidelberg}, \bibinfo{year}{2004}),
  vol.~\bibinfo{volume}{II}, pp. \bibinfo{pages}{555--730}.

\bibitem[{\citenamefont{Pfleiderer}(2009)}]{pfl09}
\bibinfo{author}{\bibfnamefont{C.}~\bibnamefont{Pfleiderer}},
  \bibinfo{journal}{Rev.\ Mod.\ Phys.} \textbf{\bibinfo{volume}{81}},
  \bibinfo{pages}{1551} (\bibinfo{year}{2009}).

\bibitem[{\citenamefont{Lee et~al.}(2006)\citenamefont{Lee, Nagaosa, and
  Wen}}]{lee06}
\bibinfo{author}{\bibfnamefont{P.~A.} \bibnamefont{Lee}},
  \bibinfo{author}{\bibfnamefont{N.}~\bibnamefont{Nagaosa}}, \bibnamefont{and}
  \bibinfo{author}{\bibfnamefont{X.-G.} \bibnamefont{Wen}},
  \bibinfo{journal}{Rev.\ Mod.\ Phys.} \textbf{\bibinfo{volume}{78}},
  \bibinfo{pages}{17} (\bibinfo{year}{2006}).

\bibitem[{\citenamefont{Kanoda}(2008)}]{kan08}
\bibinfo{author}{\bibfnamefont{K.}~\bibnamefont{Kanoda}}, in
  \emph{\bibinfo{booktitle}{The physics of organic superconductors and
  conductors}}, edited by
  \bibinfo{editor}{\bibfnamefont{A.}~\bibnamefont{Lebed}}
  (\bibinfo{publisher}{Springer-Verlag, Berlin, Heidelberg},
  \bibinfo{year}{2008}), pp. \bibinfo{pages}{623--642}.

\end{thebibliography}
\end{document}